\pdfoutput=1
\documentclass[12pt,a4paper]{article}
\usepackage{jheppub}
\usepackage[dvipsnames]{xcolor}
 
\usepackage{microtype}
\usepackage{amssymb} 
\usepackage{amsmath}
\usepackage{mathtools}
\usepackage{amsfonts}    
\usepackage{dsfont}
\usepackage{pdfpages}
\usepackage{verbatim}
\usepackage{esint}

\usepackage{tensor}
\usepackage{mathrsfs}
\usepackage{textgreek} 
\usepackage[mathscr]{euscript}
\usepackage[smalltableaux]{ytableau}
\ytableausetup{centertableaux}
\usepackage{tikz}
\usetikzlibrary{decorations.pathreplacing, decorations.markings,calc,shapes.misc,decorations.pathmorphing,patterns.meta, math, shadings, backgrounds}
\usepackage{slashed}
\usepackage{datetime}
\usepackage{subcaption}

\definecolor{MathematicaBlue}{rgb}{0.368, 0.507, 0.710}
\definecolor{MathematicaYellow}{rgb}{0.881,0.611,0.142}
\definecolor{MathematicaGreen}{rgb}{0.560,0.692,0.195}
\definecolor{MathematicaRed}{rgb}{0.923,0.386,0.209}
\definecolor{MathematicaPurple}{rgb}{0.528,0.471,0.701}
\definecolor{light-gray}{gray}{0.75}

\usepackage{hyperref}
\hypersetup{
    pdfencoding=unicode,
	colorlinks=true,
	urlcolor=Maroon,
	linkcolor=RoyalBlue,
	citecolor=Maroon,
	pdftitle={One-loop four-graviton string amplitude at finite α'},
	pdfauthor={Marco Maria Baccianti, Lorenz Eberhardt, Sebastian Mizera},
	pdfdisplaydoctitle=true,
	pdfstartview=FitH,
	linktocpage=true
}
\newcommand{\ba}{\begin{align}}

\newcommand{\be}{\begin{equation}}
\newcommand{\ee}{\end{equation}}
\def\bd{\begin{tikzpicture}}
\def\ed{\end{tikzpicture}}

\newcommand{\st}[1]{\left(\! \left( #1 \right)\! \right)}
\newcommand{\bigst}[1]{\Big(\!\!\Big( #1 \Big)\!\! \Big)}

\renewcommand\Im{\mathop{\text{Im}}}

\renewcommand\Re{\mathop{\text{Re}}}

\allowdisplaybreaks[1]
\DeclareMathAlphabet{\mathbbold}{U}{bbold}{m}{n}

\newcommand{\M}{\mathcal{M}}

\newcommand\CC{\mathbb{C}}
\newcommand\ZZ{\mathbb{Z}}
\newcommand\RR{\mathbb{R}}
\newcommand\HH{\mathbb{H}}

\newcommand\QQ{\mathbb{Q}}
\newcommand\DD{\mathbb{D}}

\newcommand\A{\mathcal{A}}
\newcommand\m{\mathsf{m}}

\renewcommand\d{\text{d}}
\newcommand\D{\mathrm{D}}
\newcommand\U{\mathrm{U}}

\newcommand{\nn}{\nonumber}
\newcommand{\Trop}{\mathrm{Trop}}
\renewcommand{\L}{\mathrm{L}}
\newcommand{\R}{\mathrm{R}}

\newcommand{\e}{\mathrm{e}}

\renewcommand{\le}{\leqslant}
\renewcommand{\ge}{\geqslant}
\renewcommand{\leq}{\leqslant}
\renewcommand{\geq}{\geqslant}

\DeclareMathOperator*{\DRes}{DRes}

\title{One-loop four-graviton\\ string amplitude at finite $\boldsymbol{\alpha'}$}

\author[1]{Marco Maria Baccianti}\emailAdd{m.m.baccianti@uva.nl}
\author[1]{\!\!, Lorenz Eberhardt}\emailAdd{l.eberhardt@uva.nl}
\author[2]{\!\!, Sebastian Mizera}\emailAdd{sm5824@columbia.edu}

\affiliation[1]{Institute for Theoretical Physics,
University of Amsterdam, Amsterdam, 1098XH, NL}
\affiliation[2]{Department of Physics,
Columbia University, New York, NY 10027, USA}

\abstract{
We evaluate the one-loop four-graviton scattering amplitude in type-II superstring theory exactly in $\alpha'$.
This result is achieved by combining physical insights into the $i\varepsilon$ prescription in string theory with a new technical application of the Rademacher expansion of modular integrals. 
We provide an implementation of our formula in $\texttt{C++}$ and use it to study the behavior of the amplitude at finite $\alpha'$ and in different kinematic regimes.
Our analysis reveals a tension between explicit computations and the saddle point analysis of Gross and Mende in the high-energy limit and suggests the presence of additional saddle points.
}

\begin{document}

\maketitle

\makeatletter
\g@addto@macro\bfseries{\boldmath}
\makeatother

\setcounter{page}{2}

\section{Introduction}
\paragraph{Motivation.} String theory provides us with UV-finite amplitudes of quantum gravity at every loop order in perturbation theory. The tree-level amplitude is the famous Virasoro--Shapiro amplitude and its properties have been widely discussed in the literature \cite{Virasoro:1969me,Shapiro:1970gy,Amati:1988tn,DiVecchia:2023frv}. 
Even though string perturbation theory is theoretically a very appealing framework, explicit computations become quickly complicated, due to the complexity of the involved integrals over the moduli space of worldsheet geometries.

In fact, comparatively little is known even about the one-loop four-graviton string amplitude. It is usually presented as an integral over the moduli space of four-punctured tori \cite{Green:1981yb,Schwarz:1982jn}, see Equation~\eqref{eq:integral-introduction} below, but this integral is very hard to perform in practice. Partial answers exist in the low-energy expansion where one encounters integrals of so-called modular graph forms \cite{Green:1999pv,Green:2008uj,DHoker:2015gmr,DHoker:2019blr,Edison:2021ebi,Huang:2024ihm,Claasen:2024ssh} and the high-energy expansion, where the integral can be done via saddle point approximation \cite{Gross:1987ar,Gross:1987kza} (although the details have never been worked out). Essentially no results are known for intermediate energies. 

\begin{figure}
\centering
\includegraphics[scale=0.48]{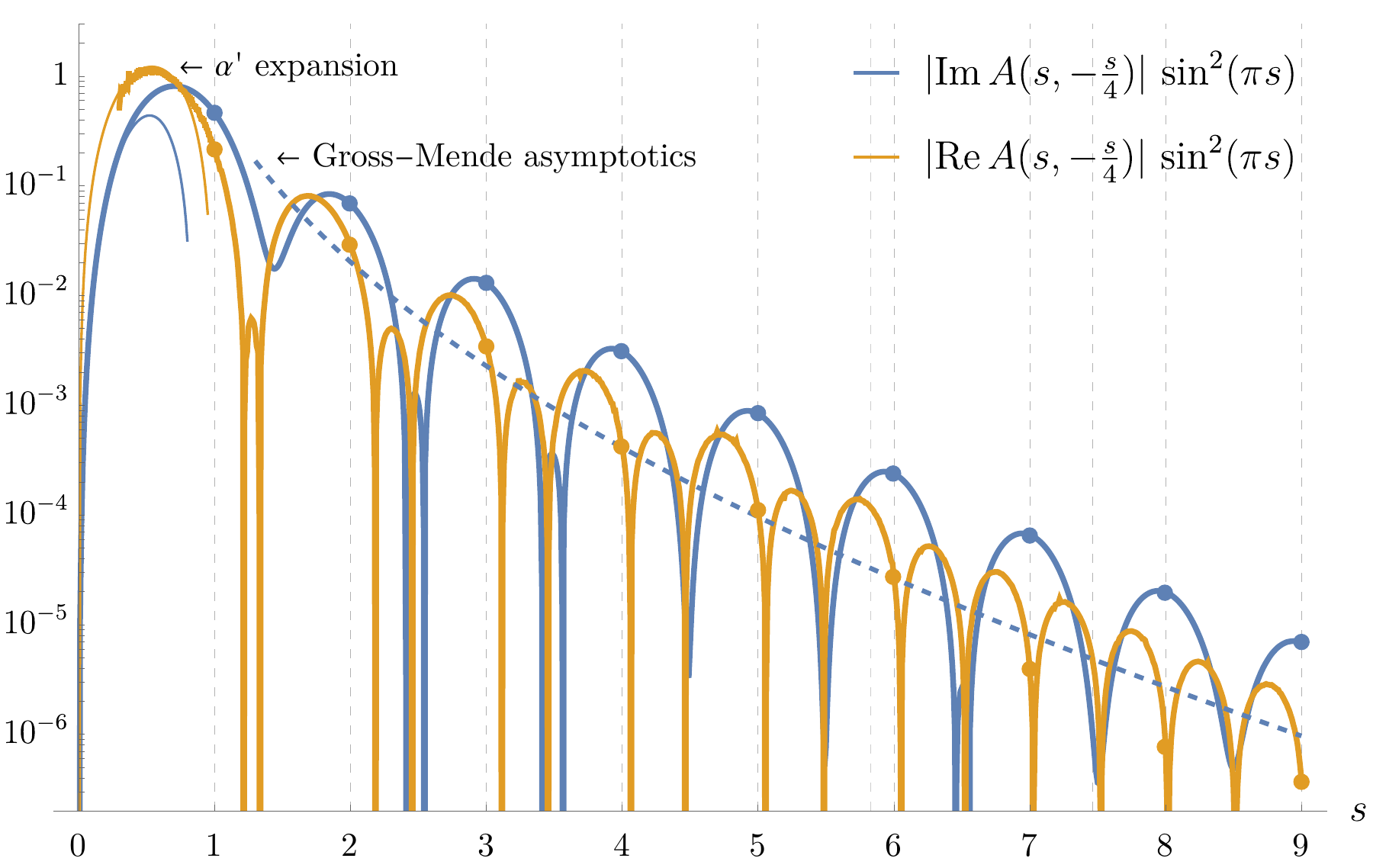}
\caption{\label{fig:intro-fixed-angle}Real (orange) and imaginary (blue) parts of the closed-string amplitude $A(s,t=-\tfrac{s}{4})$ at fixed scattering angle $\theta = 60^\circ$. The real part was obtained with \eqref{eq:intro-a-c-sum} and the imaginary part with \eqref{eq:intro-unitarity}. For readability, the functions are multiplied by a factor of $\sin^2(\pi s)$. The vertical spikes come from zeros of the real or imaginary part since we display the results on a logarithmic plot. The vertical dashed lines indicate new production thresholds opening up at $s \geq (\m_\D + \m_\U)^2$ and the dots represent integer $s$ where the amplitude can be evaluated to high precision or analytically.
The prediction from the low-energy expansion is displayed with thin lines for small $s$. The Gross--Mende prediction, which is dominated by the imaginary part, is plotted with a dashed blue curve. We find that the exponent is consistent with the numerical results, but the detailed behavior is different, indicating presence of extra saddles. For more details, see Section~\ref{sec:numerics-GM}.} 
\end{figure}

\begin{figure}
\centering
\includegraphics[scale=0.48]{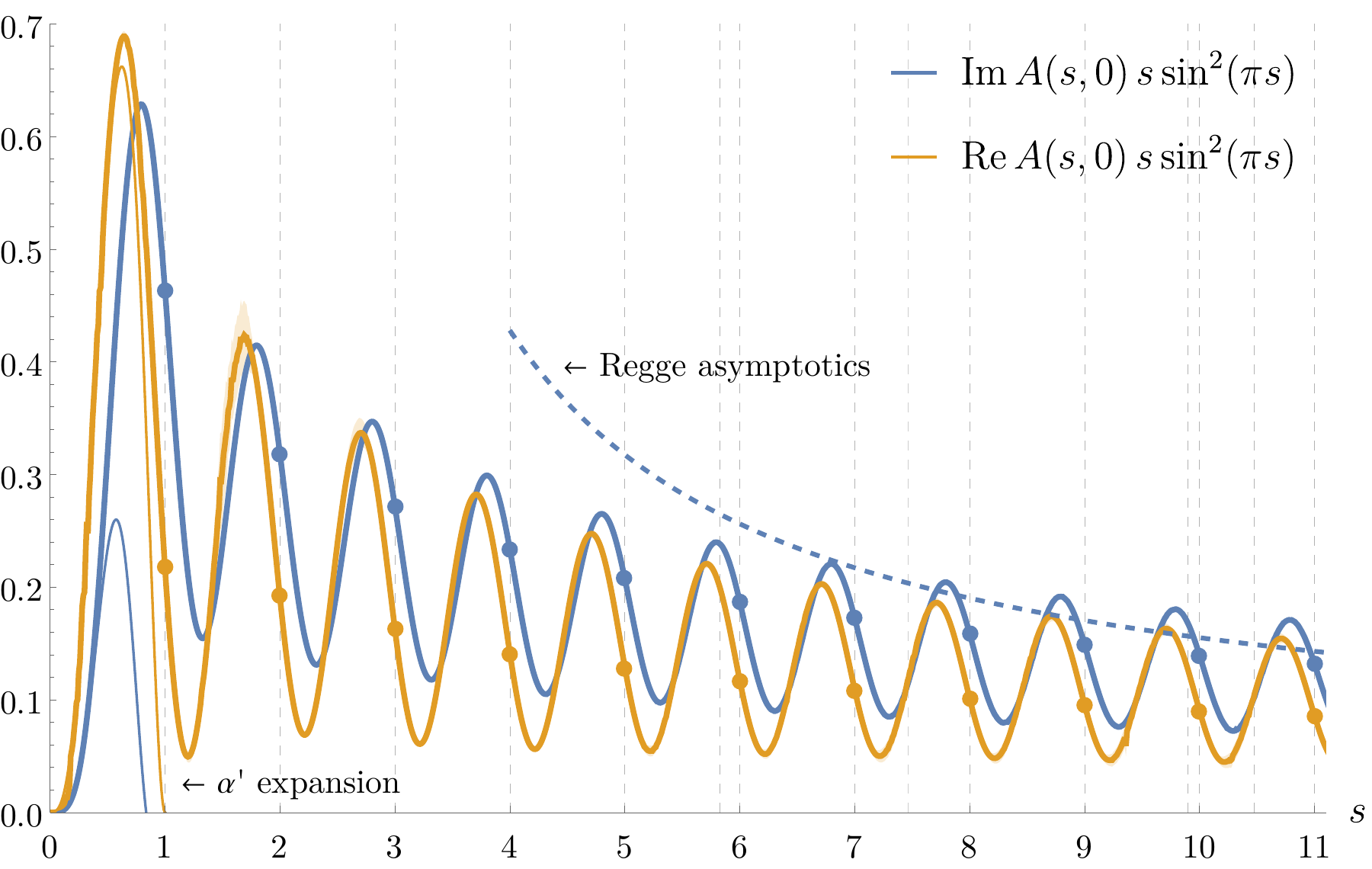}
\caption{\label{fig:intro-Regge}Real (orange) and imaginary (blue) parts of the closed-string amplitude $A(s,t=0)$ in the forward limit obtained with \eqref{eq:intro-a-c-sum} and \eqref{eq:intro-unitarity} respectively. For readability, the functions are multiplied by a factor of $s \sin^2(\pi s)$. Vertical dashed lines and dots are as in Figure~\ref{fig:intro-fixed-angle}. $95\%$ confidence intervals for the real part are given by the shaded areas, while the imaginary part is exact. The prediction from Regge analysis, which is mostly imaginary, is indicated with the dashed curve. The prediction from perturbation theory, which is a good approximation for $s \lesssim 0.4$, is shown with thin lines. For more details, see Section~\ref{sec:numerics-Regge}.}
\end{figure}

\paragraph{Main result.} The main result of this paper is a novel representation of the one-loop amplitude in type-II superstring theory. The formula takes the form of an infinite sum and is outlined below in Equation~\eqref{eq:intro-a-c-sum} and given explicitly in Equation~\eqref{eq:final Rademacher formula}. This representation features a number of sums and integrals, but they are all convergent and numerically stable. In particular, even though this formula looks arguably much more complicated than the initial integral representation, it seems to be more useful to explore the amplitude, especially at the previously inaccessible intermediate energies. To prove our point, we implemented the formula and plotted the amplitude in Figure~\ref{fig:intro-fixed-angle} at a fixed scattering angle of $60$ degrees while scanning over the energies in units where $\alpha'=4$, as well as in the forward limit in Figure~\ref{fig:intro-Regge}. The imaginary part is actually much simpler to obtain, since it can be obtained directly from unitarity methods \cite{Eberhardt:2022zay, Banerjee:2024ibt} and does not need our full formula. However, we do not know of any other way to produce the curve of the real part. With this paper, we also publish a \texttt{C++} program called \texttt{StringQMC} that implements our formula, which was used to produce Figures \ref{fig:intro-fixed-angle} and \ref{fig:intro-Regge}. We invite the reader to explore properties of the amplitude using \texttt{StringQMC} for themselves!

The infinite-sum representation is itself an amalgamation of some of the technology developed in our previous papers, especially \cite{Eberhardt:2023xck,Baccianti:2025gll}, and crucially relies on interpreting the worldsheet as a Lorentzian surface. We plan to explore its interaction with the saddle-point approximation as proposed by Gross and Mende in \cite{Gross:1987ar,Gross:1987kza} at high energies in an upcoming publication \cite{BEM}. As one can read-off from Figure~\ref{fig:intro-fixed-angle}, the saddle point approximation, which is purely imaginary, seems to produce the correct envelope of the amplitude, but it fails to reproduce more fine-grained features. Those will be explained by the presence of other relevant Lorentzian saddles \cite{BEM}.

\paragraph{A brief overview.} Let us now give a schematic overview of the origin of these results. Our starting point is the one-loop amplitude $\mathcal{A}_{g=1}$ which admits a compact formula written as an integral over the moduli space of four punctures on a torus:
\be\label{eq:integral-introduction}
\mathcal{A}_1(s,\theta) \cong t_8 \tilde t_8\int_{\mathcal{F}} \frac{\d^2 \tau}{(\Im \tau)^5} \int_{\mathbb{T}^2} \prod_{j=1}^3 \d^2 z_j \!\!\! \prod_{1 \le j<i \le 4} \e^{\alpha' p_i \cdot p_j\, G(z_i , z_j )}\, ,
\ee
where $\tau$ is the modular parameter of the torus $\mathbb{T}^2$ and is integrated over the fundamental domain $\mathcal{F}$. One of the punctures, say $z_4=0$, is fixed to the origin, while the remaining three are integrated over the torus $\mathbb{T}^2$. The rest of the integrand involves combinations of Green's functions $G(z_i , z_j )$ between each pair of punctures. The massless momenta $p_i$ are such that $s = -(p_1 + p_2)^2$ and $t = -(p_2 + p_3)^2 = \frac{s}{2}(1 - \cos \theta)$. All the polarization dependence is contained in the $t_8 \tilde t_8$ prefactor, which is always stripped off for the purposes of plots, $\mathcal{A}_1(s,t) = t_8 \tilde t_8 A(s,t)$. Overall, \eqref{eq:integral-introduction} is an $8$-dimensional integral.
There are two immediate problems with evaluating \eqref{eq:integral-introduction}, a conceptual and a practical one.

\paragraph{Conceptual problems.}
The conceptual problem is that \eqref{eq:integral-introduction} does not converge for any value of the kinematics, which is the reason why we wrote $\cong$ instead of $=$. In fact, this issue exists at any genus, including the tree-level Virasoro--Shapiro amplitude. The reason is essentially that in the neighborhood of the $s$-, $t$-, and $u$-channel degenerations, we need $\Re s <0$, $\Re t <0$, and $\Re u <0$ for convergence. But by momentum conservation we have $s + t + u = 0$, so the whole integral cannot converge simultaneously. One common approach to such a problem is to chop the integration domain in multiple pieces, evaluate each one in a separate kinematic channel, analytically continue back to common kinematics, and then sum all the contributions (see, e.g., \cite{DHoker:1993hvl,DHoker:1993vpp,DHoker:1994gnm}). This approach does not work numerically because analytic continuation is unstable. A more natural and physical solution is to devise a modified integration contour that would allow us to make sense of \eqref{eq:integral-introduction} directly in physical kinematics. See \cite{Berera:1992tm,Witten:2013pra,Eberhardt:2022zay} for previous discussions. 

\paragraph{Practical problems.}
After resolving the conceptual issues by modifying the integration contour appropriately, the practical problem with evaluating \eqref{eq:integral-introduction} is that the integrand oscillates rapidly close to the boundaries of moduli space.
This feature is the ``sign problem'' of string amplitudes and makes it near impossible to compute \eqref{eq:integral-introduction} directly, at least for high energies. As an aside, let us mention that a direct integration strategy was successfully carried out for higher-point tree-level amplitudes with closed \cite{Eberhardt:2024twy} and open strings \cite{Eberhardt:2024twy, Figueiredo:2025fnr}, but it rapidly becomes numerically unstable for high energies where one gets large cancellations in the integral. 

\paragraph{Solutions.}
The conceptual problem of defining the physical moduli space for string amplitudes has been a subject of ongoing investigations \cite{Eberhardt:2022zay,Eberhardt:2023xck,Banerjee:2024ibt,Manschot:2024prc,Baccianti:2025gll}. In particular, with Chandra and Hartman \cite{Baccianti:2025gll} we introduced a physical prescription for evaluating integrals over the fundamental domain $\mathcal{F}$. It uses the \emph{double Rademacher contour} that extends the classic work of Rademacher in analytic number theory \cite{Rademacher, RademacherZuckerman} to modular integrals.

We apply the double Rademacher contour to $\mathcal{A}_1$. Extra work is needed to deal with the infinite number of branch cuts stemming from $G(z_i, z_j)$ in \eqref{eq:integral-introduction}. We go over these steps in Section~\ref{sec:Rademacher expansion}. Together, they solve also the practical issues with the sign problem.

The net result is an expression for $\mathcal{A}_1$ as an infinite sum of the form
\be\label{eq:intro-a-c-sum}
\mathcal{A}_1 = t_8 \tilde t_8 \sum_{c=1}^\infty \sum_{\begin{subarray}{c} a=0 \\ \text{gcd}(a,c)=1 \end{subarray}}^{c-1} \sum_{n_\L,n_\D,n_\R=0}^{c-1} \mathrm{e}^{i (\mathrm{phase})} A_{a/c}^{n_\L,n_\D,n_\R}\ ,
\ee
where an explicit formula for the phase and $A_{a/c}^{n_\L,n_\D,n_\R}$ will be given in Equation~\eqref{eq:A as sum over windings} and \eqref{eq:final Rademacher formula}. The summation indices have a certain interpretation in terms of windings around the torus. \eqref{eq:intro-a-c-sum} is not an absolutely convergent sum, but due to the presence of the phases, it still seems to conditionally converge fairly rapidly.
Thus in practical calculations one can truncate the sum over $c$ and expect a good numerical approximation of the amplitude. 

\paragraph{Special cases and cross checks.}
We made a number of cross checks in order to verify the validity of the expansion \eqref{eq:intro-a-c-sum}. The strongest one comes from unitarity cuts \cite{Aoki:1990yn,Berera:1992tm,Pius:2018crk,Eberhardt:2022zay}, which compute the imaginary part of the amplitude $\A_1$ and schematically takes the form
\begin{align}\label{eq:intro-unitarity}
\Im \A_1^{(hh \to hh)} = \sum_{\D,\U} \int &\d^{10} \ell_\D\, \d^{10}\ell_\U\; \A_0^{(hh \to \D \U)} {\A_0^\ast}^{(hh \to \D \U)} \\
&\times \delta^+(\ell_\D^2 - \m_\D^2)\, \delta^+(\ell_\U^2 - \m_\U^2) \, \delta^{10}(\ell_\D + \ell_\U - p_1 - p_2)\, .\nonumber
\end{align}
On the right-hand side, the one-loop scattering process $hh \to hh$ is constructed by gluing together tree-level amplitudes $\A_0$ with $hh \to \D \U$ for all kinematically allowed intermediate states $\D$ and $\U$ with momenta $\ell_\D$, $\ell_\U$ masses $\m_\D$, $\m_\U$ respectively.
Indeed, the most difficult part of this computation is the sum over the intermediate states $\D$ and $\U$ that flow through the unitarity cut. From the space-time perspective, such sums are near-impossible to perform ``by hand'' since string theory contains an exponential number of states at every mass level, each having its own polarization structure.

Instead, in our approach, the formula \eqref{eq:intro-unitarity} is derived directly from the worldsheet and all the complicated state sums are summarized in certain explicit polynomials integrated over the phase space \cite{Eberhardt:2022zay,Banerjee:2024ibt}.
This is due to a remarkable simplification in the computation of \eqref{eq:intro-unitarity}, which is an incarnation of the \emph{double copy relations} \cite{Kawai:1985xq,Bern:2008qj}. It essentially means that all the results can be imported from the type-I open-string scattering \cite{Eberhardt:2022zay}. The resulting expression, in Baikov parametrization \cite{Baikov:1996iu} was previously derived in \cite{Banerjee:2024ibt} and is given in Equation~\eqref{eq:imaginary part Baikov representation}.
Hence, by comparison, the imaginary part $\Im \A_1$ is rather trivial to evaluate compared to the full $\A_1$. Checking that $\Im \A_1$ computed using these two entirely different expressions is in numerical agreement provides a strong check that the whole $\A_1$ is correct.

We also discuss various special cases, such as the one-loop decay widths and mass shifts of the string spectrum which can be read off from the amplitude and compare our formula to the analogous open string formula obtained in \cite{Eberhardt:2023xck}.

\paragraph{Limitations.} Our formula has also a number of limitations. It is, perhaps surprisingly, fairly inefficient at reproducing the low-energy behavior of the amplitude. The physical reason for this is that the amplitude has a massless branch cut, which manifests itself in a failure of the sum over $c$ in \eqref{eq:intro-a-c-sum} to converge for low energies. One can see this in Figure~\ref{fig:intro-fixed-angle}, where the numerical error becomes significant for low energies, $s \lesssim 1$. For such low energies, it is more practical to use the low-energy expansion methods that have been developed in the literature or direct numerical integration. 

The complexity of the formula also increases for higher energies, since one needs to account for more and more intermediate states. In practice, we have been able to compute the full amplitude to about $s\sim 12$ with a $\sim 5\%$ error, while partial answers such as the mass shifts can be computed much more precisely, with errors of $ \sim 0.0001 \%$ up to $s \sim 35$ (on a regular computer).
\paragraph{Outline.}
This paper is organized as follows. We start by showing some results obtained from our explicit formula in Section~\ref{sec:numerics}. The rest of the paper is devoted to deriving it and exploring its properties analytically. We explain the step-by-step derivation in Section~\ref{sec:Rademacher expansion}. In Section~\ref{sec:special cases}, we explain various simplifications that occur in the forward limit, and for the mass shifts. We also make contact with the imaginary part as computed from the Baikov representation and the low energy expansion that has previously been discussed in the literature. Conclusions are presented in Section~\ref{sec:conclusions} and some important identities involving Bessel functions in Appendix~\ref{app:Bessel function identities}.

\section{\label{sec:numerics}Numerical implementation and results}

Attached to this paper, we provide a numerical implementation of the Rademacher formula \eqref{eq:final Rademacher formula} (including the forward limit \eqref{eq:final Rademacher formula forward limit}) and the Baikov formula \eqref{eq:imaginary part Baikov representation} in \texttt{C++}. It allows one to reproduce all the results presented in this work.

The \texttt{C++} program called \texttt{StringQMC} uses the quasi Monte--Carlo integrator \cite{Borowka:2018goh} developed for high-precision computations of Feynman integrals. While the full documentation can be found in the source code of the program, the basic options of \texttt{StringQMC} are:
\begin{align}
&\texttt{
StringQMC -cMin 1 -cMax 10}\nn\\
&\qquad\qquad\quad\;\;\;\texttt{-sMin 0.01 -sMax 10.00 -sStep 0.01}\nn\\
&\qquad\qquad\quad\;\;\;\texttt{-thetaMin 60 -thetaMax 60 -thetaStep 1}\nn\\
&\qquad\qquad\quad\;\;\;\texttt{-epsrel 1e-2 -epsabs 1e-8}\nn
\end{align}
The first 8 options specify the range of $(c,s,\theta)$ computed. The final 2 options specify the relative and absolute error requested. Whenever $\theta=0$ occurs, the program automatically switches to using the forward-limit formula \eqref{eq:final Rademacher formula forward limit}. Optionally, one can also specify \texttt{-mode "Baikov"} to use the Baikov expression \eqref{eq:imaginary part Baikov representation}. We invite readers to consult the source code for more options and the format of the output.

To exemplify the type of results that can be obtained, we focus on fixed-angle scattering at $\theta = 60^\circ$ and forward scattering at $\theta = 0^\circ$. Let us discuss them in turn.

\subsection{\label{sec:numerics-GM}Fixed-angle scattering}

The main result in fixed-angle scattering is the plot in Figure~\ref{fig:intro-fixed-angle}. It displays the real and imaginary part of the amplitude $A(s,t)$ with polarization dependence stripped away, $\mathcal{A}_1 = t_8 \tilde t_8 A$, plotted up to $s \leq 9$ in the units of $\alpha' = 4$, in which the string spectrum has the simple form $m^2 \in \mathbb{Z}_{\ge 0}$. We use the Rademacher formula for the real part and the Baikov formula for the imaginary part.

Recall that the amplitude has a series of double poles at positive integers. Therefore, for readability, we multiply the plotted functions by $\sin^2 (\pi s)$. (If the amplitude could be genus-resummed, it would result in the poles moving to the lower half-plane of $s$, which would similarly produce a smooth plot.) At integer values of $s$, the resulting quantity therefore coincides with the values of the mass shifts and decay widths discussed in Section~\ref{sec:mass shifts}, up to a factor of $\pi^2$. We evaluate them at high precision with $c_{\mathrm{max}} = 100$ and plot them as dots in Figure~\ref{fig:intro-fixed-angle}.

Since computational complexity grows with $s$, at non-integer $s$ we use $c_\text{max} = (40,20,10,6,3,1)$ for $s < (1, 2, 5, 6, 8, 9)$ respectively. No extrapolation to $c_{\mathrm{max}} \to \infty$ is used.\footnote{Such an extrapolation does not seem to be particularly useful. The error $A-\sum_{c=1}^{c_\text{max}} A_c$, where $A_c$ is the result in \eqref{eq:intro-a-c-sum} does not seem to have an asymptotic expansion in $c_\text{max}^{-1}$, because $A_c$ crucially depends on the number-theoretic properties of $c$, for example because of the appearance of $\text{gcd}(a,c)$ in \eqref{eq:intro-a-c-sum}. Thus standard convergence-accelerating transforms are not applicable.} These cutoffs are enough to recognize broad features of the amplitude, such as exponential suppression, but we do not expect that finer details, such as positions of zeros are accurate for larger values of $s$.

In Figure~\ref{fig:intro-fixed-angle}, we plot the low-energy approximation from Section~\ref{sec:low energy expansion}, which summarizes all perturbative results about this amplitude known to date. As expected, they approximate $A(s,t)$ relatively well up to $s \lesssim 0.3$. The perturbative expansion has to break down at the radius of convergence $|s|=1$.

\subsection{Comparison with Gross--Mende}
In Figure~\ref{fig:intro-fixed-angle}, we also plot the Gross--Mende behavior. It arises by evaluating the moduli space integral via saddle point approximation, which becomes a good approximation at large $s$, where the integral is sharply peaked around the saddle point. Gross and Mende made a conjecture about the leading saddle, which is a double cover of a four-punctured sphere, branched over the insertions of the sphere \cite{Gross:1987kza, Gross:1987ar}. Because this is a two-sheeted cover, the on-shell action is half of the tree-level value
\begin{align}
S &\equiv s \log s+t \log(-t)+u \log(-u)\\
&= - s \bigg[ \sin^2(\tfrac{\theta}{2}) \log \left( \sin^2(\tfrac{\theta}{2}) \right) + \cos^2(\tfrac{\theta}{2}) \log \left( \cos^2(\tfrac{\theta}{2}) \right) \bigg]\, ,
\end{align}
where $\cos \theta = 1 + \frac{2t}{s}$ and $s+ t + u = 0$.
This leads to the general prediction that the amplitude should asymptotically behave as
\be 
A_\text{GM}(s,t)=i\, F(\tfrac{t}{s})\, s^{-4} \mathrm{e}^{-\frac{1}{2}S}\ .\label{eq:one-loop amplitude saddle point contribution}
\ee
The factor $s^{-4}$ is a one-loop effect around the saddle and arises because all entries of the Hessian are proportional to $s$ and we are computing an 8-dimensional integral and thus $\sqrt{\det \text{Hess}} \propto s^4$.
The function $F$ capturing the angular dependence of the amplitude can in principle be evaluated by computing the determinant of the Hessian around the saddle point. The determinant of the Hessian turns out to be negative, and thus \eqref{eq:one-loop amplitude saddle point contribution} is purely imaginary, which we indicated by including an overall imaginary unit. Since $F$ is a constant for a fixed angle, the specific function plotted using a dashed curve in Figure~\ref{fig:intro-fixed-angle} is $i s^{-4} \e^{-\frac{1}{2}S}$.

The exponential behavior seems to match, but the finer details do not. In \cite{BEM}, we will explain this difference through presence of new Lorentzian saddles beyond the one described by Gross and Mende. They have the same real part of the saddle point value, but also have an imaginary part that leads to the oscillating contributions present in Figure~\ref{fig:intro-fixed-angle}. 

\begin{figure}
\centering
\includegraphics[width=\textwidth]{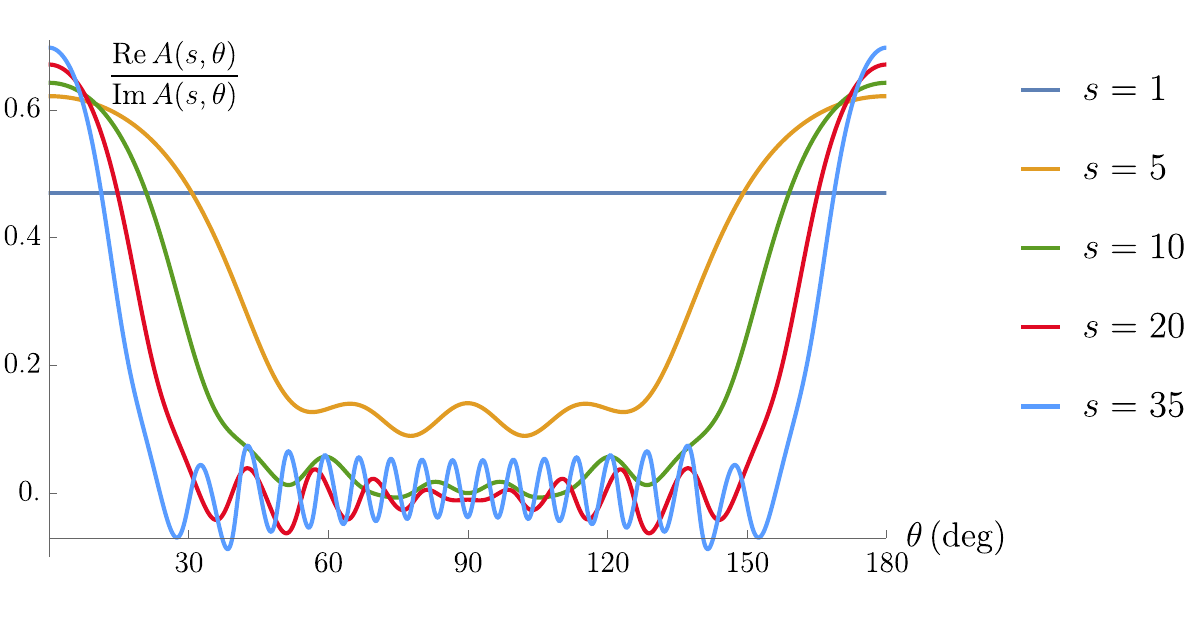}
\caption{\label{fig:reim}Ratio of the real to the imaginary part of the amplitude $A(s, t= \frac{s}{2}(1-\cos\theta))$ plotted as a function of the scattering angle $\theta$ for different values of $s$.}
\end{figure}

\begin{figure}
\centering
\includegraphics[scale=0.7]{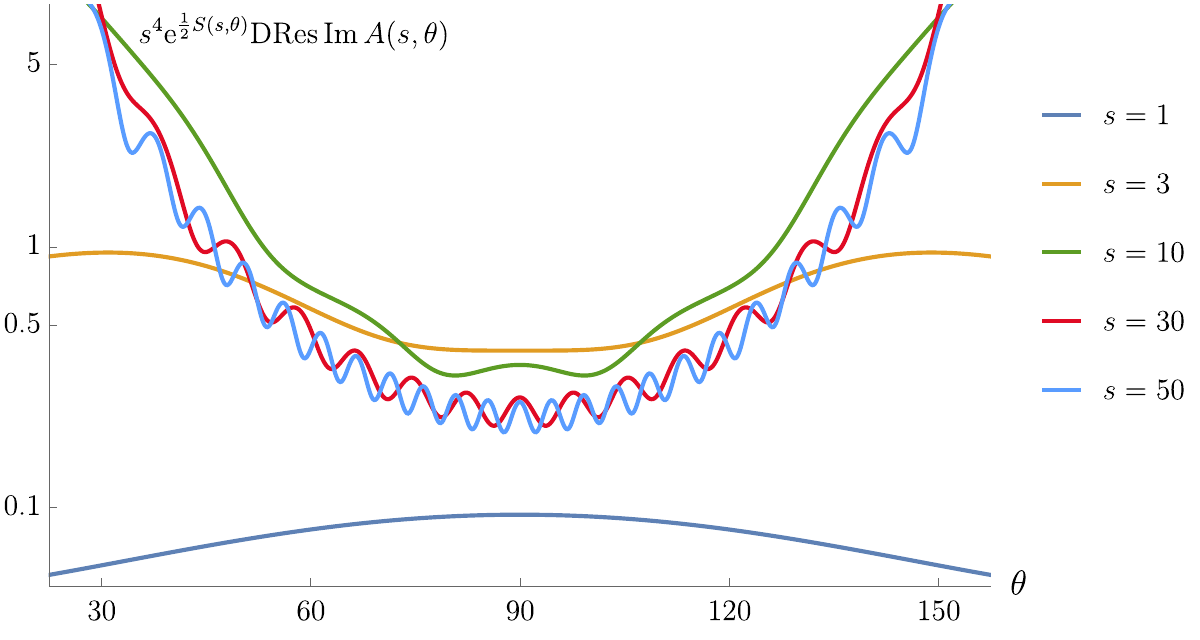}
\caption{\label{fig:imGM}The imaginary part of the amplitude $A(s, t= \frac{s}{2}(1-\cos\theta))$ normalized by the general Gross--Mende behavior \eqref{eq:one-loop amplitude saddle point contribution}. We observe that at large $s$ the ratio stabilizes, up to small oscillations.}
\end{figure}

To further study the validity of the Gross--Mende approximation, we plot the angular dependence in Figure~\ref{fig:reim} and \ref{fig:imGM}. This is simplest to do for integer $s$, where the amplitude has double poles whose coefficients encode the mass shifts of massive string states \cite{Eberhardt:2022zay}.

First, the prefactor of the saddle point \eqref{eq:one-loop amplitude saddle point contribution} is purely imaginary. This predicts that at higher and higher values of $s$, the imaginary part should dominate. (A similar prediction in the forward limit comes from Regge theory and crossing symmetry \cite{Banerjee:2024ibt}). To test this prediction, we plot the ratio of the real and imaginary part of $A(s,t = \tfrac{s}{2}(1-\cos \theta))$ as a function of $\theta$ for different values of $s$. We find that the imaginary part is indeed bigger, but not parametrically so. The oscillations in Figure~\ref{fig:reim} do not decay with larger values of $s$. As will be described in \cite{BEM}, the culprit are again the presence of further Lorentzian saddles that give additional contributions to \eqref{eq:one-loop amplitude saddle point contribution} which are not purely imaginary. Based on the results of \cite{Banerjee:2024ibt}, we also expect that the imaginary part dominates in the strict forward limit ($\theta = 0^\circ$ and $180^\circ$), except this would not be apparent until extremely high energies due to a logarithmic behavior in $s$.

Next, we test the asymptotic behavior $i s^{-4} \mathrm{e}^{-\frac{1}{2}S(s,\theta)}$ by normalizing the decay widths by it. We see from Figure~\ref{fig:imGM} that up to the oscillations mentioned before, the result nicely stabilizes for large $s$, therefore giving strong support to the validity of saddle point approximation. 

\subsection{\label{sec:numerics-Regge}Forward scattering}

The main plot illustrating forward scattering amplitude, $t = \theta = 0$, is given in Figure~\ref{fig:intro-Regge} for $s<11$. For readability, the plot is normalized by the factor of $s \sin^2(\pi s)$. In this case, the advantage is that we can use simplified expression for the scattering amplitude \eqref{eq:final Rademacher formula forward limit} and hence achieve higher accuracy. We use $c_{\mathrm{max}} = 40$ for $s < 1$ and $c_\mathrm{max} = 10$ for $s < 12$. Other conventions for Figure~\ref{fig:intro-Regge} and the same as in Figure~\ref{fig:intro-fixed-angle}.

\begin{figure}
\centering
\includegraphics[scale=0.24]{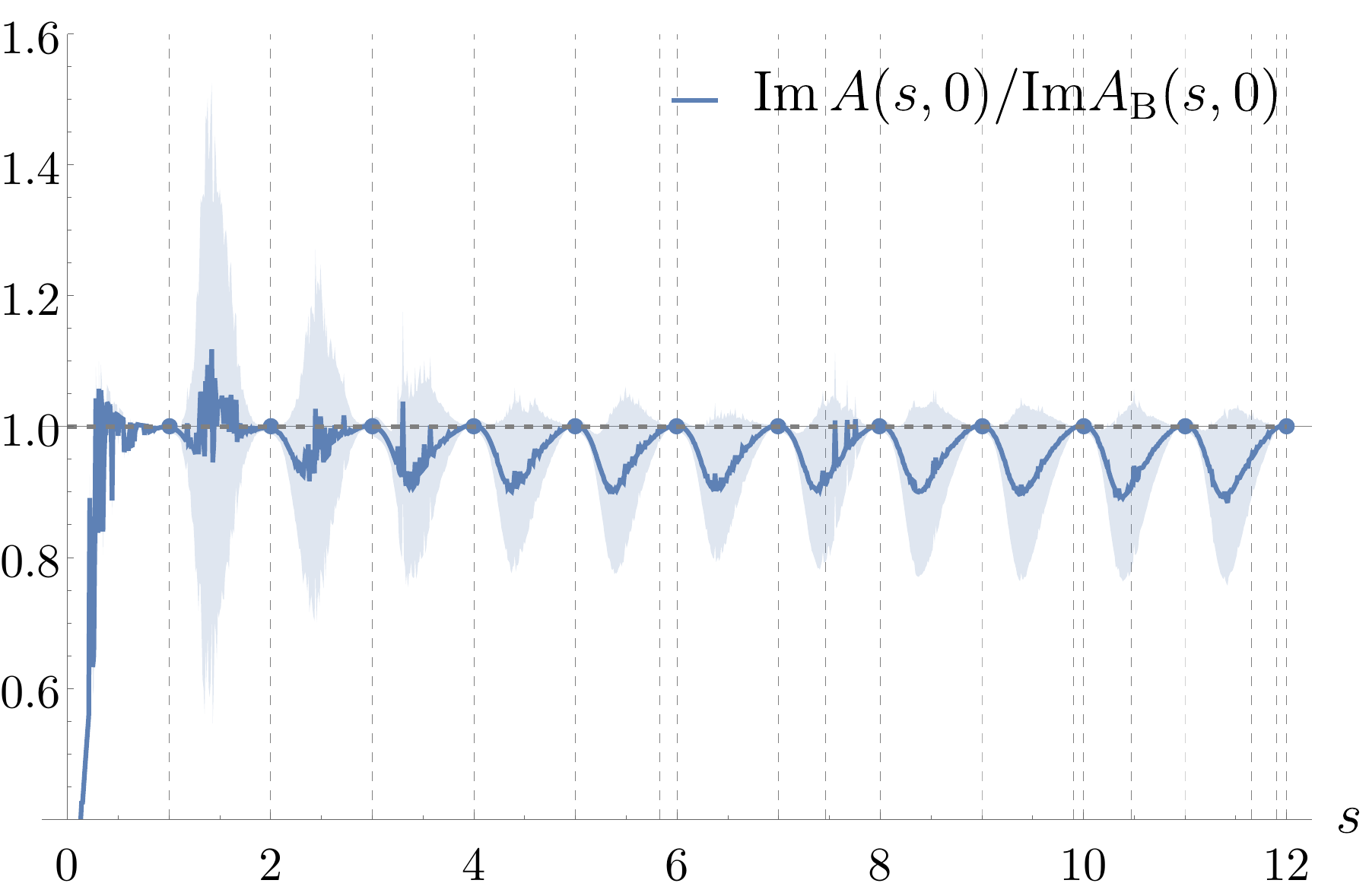}\hspace{-1em}
\includegraphics[scale=0.52]{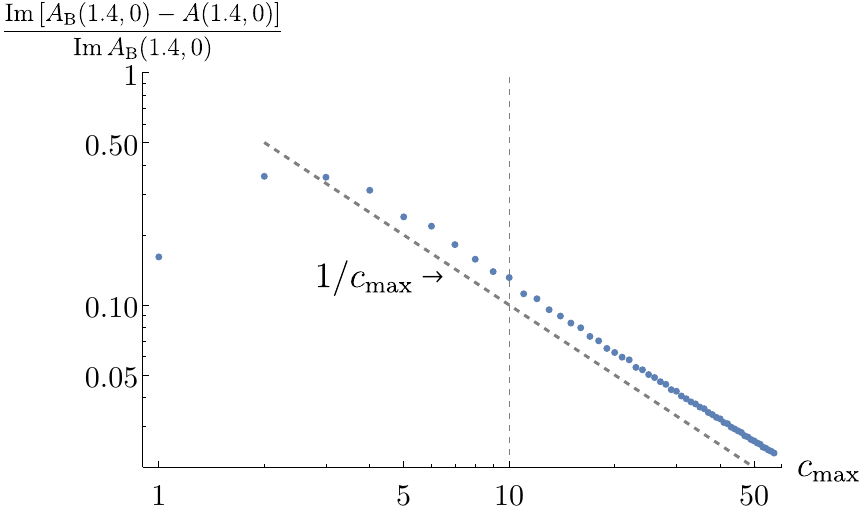}
\caption{\label{fig:Baikov-comparison}\textbf{Left:} The ratio between the imaginary part of the amplitude computed using the Rademacher and Baikov formulae. The former is truncated at $c_{\text{max}} = 10$ and extrapolated (for $s > 1$), while the latter can be treated as being exact. 95\% confidence intervals plotted as shaded areas come from extrapolation errors. Monte Carlo sampling errors are kept at $< 1\%$ and hence are negligible in comparison. \textbf{Right:} The relative error of the Rademacher expansion at $s=1.4$, which is one of the points with poorest convergence, as a function of $c_{\mathrm{max}}$. Around $c_{\mathrm{max}}=10$, which is the cutoff used in most of the results, the Rademacher series reaches around $13\%$ precision.}
\end{figure}

For each value of $s$, we extrapolate to $c_{\mathrm{max}} \to \infty$ by fitting the data to the ansatz $\alpha + \beta c_{\text{m}}^{\gamma}$ for $c_{\text{m}} \in \{ 5,6,\ldots,c_\text{max}\}$. Provided $\gamma < 0$, we then plot the central value for $\alpha$ and the associated $95\%$ confidence intervals. We can test the quality of this procedure by comparing the result of the extrapolation of the imaginary part to the Baikov result, which is essentially exact. This is done in Figure~\ref{fig:Baikov-comparison} (left). We find that even $c_{\mathrm{max}} = 10$ gives around $10\%$ accuracy for the central value and good agreement within the error bars. As explained in Section~\ref{sec:low energy expansion}, the Rademacher formula has poor convergence as $s \to 0$, which is visible in the plot. As an example of how increasing $c_{\mathrm{max}}$ can improve the results, in Figure~\ref{fig:Baikov-comparison} (right) we give a more detailed comparison for $s = 1.4$, which is one of the points with largest error bars. We find that pushing $c_{\mathrm{max}}$ to higher values indeed results in a more accurate prediction.
We provide further evidence for convergence of the Rademacher formula in Section~\ref{sec:mass shifts}.

\subsection{Comparison with Regge}

The reason why we included an additional normalization by $s$ in Figure~\ref{fig:intro-Regge} is that Regge physics predicts that in the $s \to \infty$, $t=0$ limit, the one-loop amplitude behaves to leading order as as $\A_1 \sim s^3$, see \cite{Banerjee:2024ibt}. After stripping away the polarization prefactors, $t_8 \tilde t_8 \sim s^4$, Regge prediction gives $A(s,t=0) \sim s^{-1}$ \cite{SUNDBORG1988545,Amati:1987wq,Amati:1987uf}. Including prefactors and logarithmic corrections, one in fact has \cite{Banerjee:2024ibt}
\be
A(s,t=0) \sim A_{\mathrm{R}}(s,0) \equiv \frac{i \pi^2}{12 s \log^2 s}\, .
\ee
This prediction is plotted with a dashed line in Figure~\ref{fig:intro-Regge}. It does not fit particularly well since subleading corrections are only suppressed as $(\log s)^{-1}$.

A more direct comparison is given in Figure~\ref{fig:Baikov-comparison}, where we summarize the ratios of the imaginary part computed using the Baikov formula with the Gross--Mende and Regge predictions. Up to oscillations, the ratio approaches $1$ in the Regge case. From \cite{Banerjee:2024ibt}, we know that the oscillations die out at extremely large values of $s$. As mentioned earlier, we do not expect the same to be true for the Gross-Mende ratio.

\begin{figure}
\centering
\includegraphics[scale=0.48]{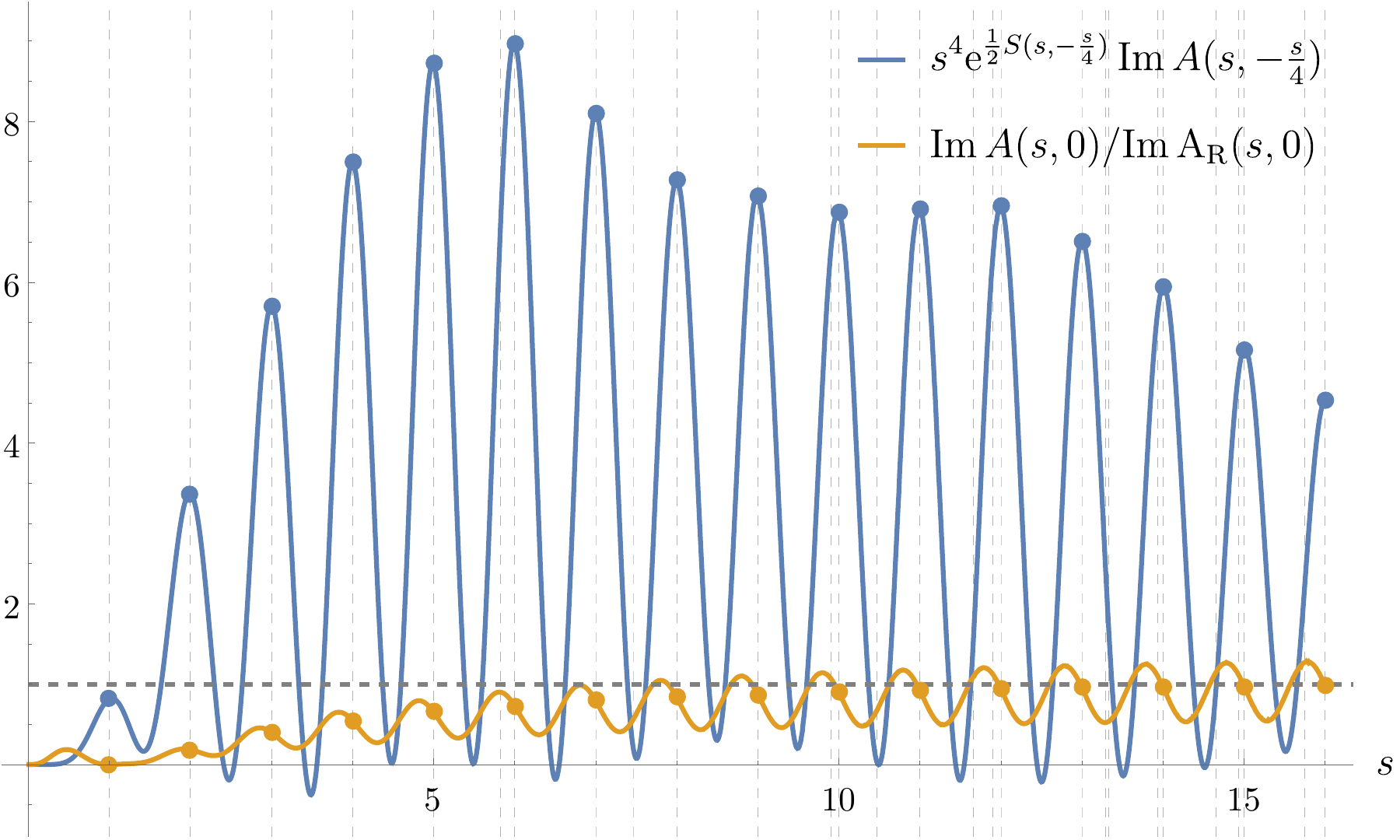}
\caption{\label{fig:Baikov-ratios}Ratios between the imaginary part of the amplitude computed using the Baikov formula and the asymptotic prediction: in the fixed-angle limit (blue) and in the Regge limit (orange).}
\end{figure}

\section{Rademacher expansion of the one-loop amplitude}\label{sec:Rademacher expansion}
In this section, we derive a closed form expression for the one-loop amplitude via the Rademacher expansion. The final result is given by \eqref{eq:A as sum over windings} and \eqref{eq:final Rademacher formula}. The final formula is quite complicated, but explicit and can be readily implemented. We analyze various special cases in Section~\ref{sec:special cases}.

\subsection{Starting point}
The one-loop amplitude in type IIA or type IIB string theory is given by the following formal integral over moduli space \cite{Green:1981yb, Schwarz:1982jn}:
\be
A \cong \int_{\mathcal{F}} \frac{\d^2 \tau}{(\Im \tau)^5} \int_{\mathbb{T}^2} \prod_{j=1}^3 \d^2 z_j\ \prod_{1 \le j<i \le 4} |\vartheta_1(z_{ij} | \tau)|^{-2s_{ij}}\, \mathrm{e}^{\frac{2\pi s_{ij} (\Im z_{ij})^2}{\Im \tau}}\ . \label{eq:4-point function graviton scattering type II}
\ee
From now on, we work in the units where $\alpha'=4$ and mostly-plus signature. We already stripped off the polarization dependence $t_8 \tilde{t}_8$, the dependence on the string coupling, as well as the momentum-conserving delta-function.
The Mandelstam variables $s_{ij} =- (p_i + p_j)^2$ are defined as
\be 
s=s_{12}\ , \quad t=s_{14}\ , \quad u=s_{13}\ .
\ee
We also have $s_{12}=s_{21}=s_{34}=s_{43}$ and similarly for the other invariants. To obtain \eqref{eq:4-point function graviton scattering type II} one already had to integrate over the fermionic directions in moduli space. This can be done in a canonical way for one-loop amplitudes. Thus the remaining integral only runs over the bosonic moduli space of surfaces.

\begin{figure}
    \centering
    \begin{tikzpicture}[scale=.7]
        \draw[domain=0:1800, variable=\t,samples=200, very thick, smooth] plot ({(1-.0003*\t)*cos(\t)},{(1-.0003*\t)*sin(\t)}); 
        \draw[domain=0:1800, variable=\t,samples=200, very thick, smooth] plot ({2+(1-.0003*\t)*cos(\t)},{4+(1-.0003*\t)*sin(\t)}); 
        \draw[dashed, very thick, bend right=10] (-.65,-1) node[below] {$\mathscr{D}_1$} to (2.4,5);
        \draw[very thick, bend right=10] (1,0) to (3,4);
        \draw[very thick, bend right=10] (-.95,0) to (1.05,4);
        \draw[very thick, bend right=10] (1,0) to (6,1);
        \draw[very thick, bend right=10] (3,4) to (5,5);
        \draw[very thick, bend right=10] (6,1) to (5,5);
        \draw[very thick, bend right=10] (7.95,1) to (6.95,5);
        \draw[domain=0:1800, variable=\t,samples=200, very thick, smooth] plot ({6-(1-.0003*\t)*cos(\t)},{5-(1-.0003*\t)*sin(\t)}); 
        \draw[domain=0:1800, variable=\t,samples=200, very thick, smooth] plot ({7-(1-.0003*\t)*cos(\t)},{1-(1-.0003*\t)*sin(\t)}); 
        \draw[very thick, bend right=10, dashed] (7.1,-.5) node[below] {$\mathscr{D}_2$} to (5.5,6.3);
    \end{tikzpicture}
    \caption{The Lorentzian integration contour near two boundary divisors $\mathscr{D}_1$ and $\mathscr{D}_2$.}
    \label{fig:Lorentzian contour}
\end{figure}
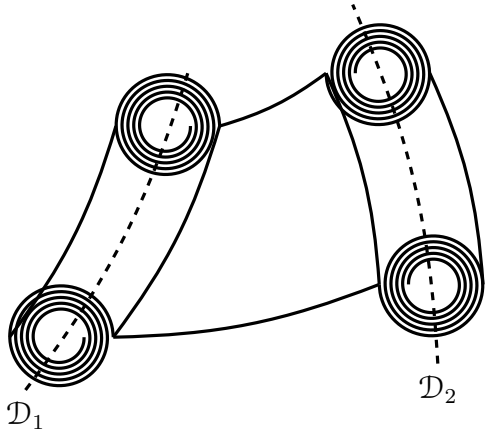

To make the definition of the integral \eqref{eq:4-point function graviton scattering type II} complete, we have to specify the precise contour over which we integrate. The integral is potentially divergent as $\Im \tau \to \infty$, as well as for $z_{ij} \to 0$, depending on the kinematics. Such boundary divisors represent different IR-regions in moduli space and manifest themselves in terms of analytic features in the amplitude. One way to properly define the integral is to modify the integration contour and make use of the string-theoretic $i \varepsilon$ prescription \cite{Witten:2013pra}. This specifies a particular integration contour $\Gamma \subset \mathcal{M}_{1,4}^{\mathbb{C}}$. The integration contour never reaches these boundary divisors $\mathscr{D}_i$ in the moduli space, but instead starts winding around them in a Lorentzian direction, see Figure~\ref{fig:Lorentzian contour} for a schematic representation. In particular the question of compactifying the moduli space $\mathcal{M}_{1,4}$ does not arise in this Lorentzian formulation. 
While this procedure sounds quite complicated, most of the singularities are actually very benign. In particular, the divergences for $z_{ij} \to 0$ lead to the resonance poles of the amplitude and in particular do not introduce branch cuts. Thus there is no need to carry the deformation of the contour for these degenerations explicitly in the computation, though in principle the contour is understood at any multiplicity \cite{Eberhardt:2024twy}. The branch cuts come exclusively from the region $\Im \tau \to \infty$ in moduli space and there usage of the $i \varepsilon$ prescription is particularly important. The relevant contour is explained in detail in \cite{Baccianti:2025gll}. It deforms the integral over $\Im \tau$ away from the real axis and let it run along an imaginary direction. Thanks to the term $(\Im \tau)^{-5}$ and the oscillating nature of the rest of the integrand in that Lorentzian direction of moduli space, this gives a convergent definition of the integral. 

We will be mostly concerned with the evaluation of the real part of the amplitude. The imaginary part is comparatively simple since it originates from a small neighborhood of $\Im \tau \to \infty$ and can be extracted by a local expansion of the integrand \cite{Eberhardt:2022zay,Banerjee:2024ibt}. The result takes the form of a sum over unitarity cuts, see Equation~\eqref{eq:imaginary part Baikov representation}.
Thus, we will compute the real part of the amplitude, for which we will take the average of the integral running along the positive imaginary direction and the negative imaginary direction. This contour was denoted by $\fint_{\mathcal{F}}$ in \cite{Baccianti:2025gll}. Thus, we will compute the integral
\be 
\Re A= \fint_{\mathcal{F}} \frac{\d^2 \tau}{(\Im \tau)^5} \int_{\mathbb{T}^2} \prod_{j=1}^3 \d^2 z_j\ \prod_{1 \le i<j \le 4} |\vartheta_1(z_{ij} | \tau)|^{-2s_{ij}}\, \mathrm{e}^{\frac{2\pi s_{ij} (\Im z_{ij})^2}{\Im \tau}}\ . \label{eq:4-point function graviton scattering type II regularized}
\ee
As mentioned above, the imaginary part $\Im A$ admits a much simpler formula which will be given in \eqref{eq:imaginary part Baikov representation}.

\subsection{Complexification and Rademacher contour}
We want to apply the Rademacher evaluation of modular integrals  developed in \cite{Baccianti:2025gll} to evaluate this integral as explicitly as possible. The Rademacher procedure evaluates the integral in closed form in terms of the local behavior of the integrand near the Lorentzian singularities in the complexification $\mathcal{M}_{1,4}^\CC$ of moduli space. This will means that some of the integrands in \eqref{eq:4-point function graviton scattering type II} become simple to perform. The general
The Rademacher formula tells us that
\begin{align} 
    \fint_{\mathcal{F}} \d^2 \tau\, f(\tau,\tilde{\tau})=\sum_{c=1}^\infty \sum_{\begin{subarray}{c} a=0 \\ (a,c)=1 \end{subarray}}^{c-1} \int_{ \longrightarrow}\!\!\!\d \tau\int_{C_{a/c}}\!\!\!\d \tilde{\tau} \,  \bigg[\frac{1}{12i}\Big(\tau-\tilde{\tau}+\frac{2a}{c}\Big)+i s(a,c)\bigg]\, f(\tau,\tilde{\tau})\label{eq:general Rademacher formula}
\end{align}
for a modular covariant integrand $f(\tau,\tilde{\tau})$. We denote $\tilde{\tau}=-\bar{\tau}$, so that $\tilde{\tau} \in \HH$. Here, $C_{a/c}$ is the Ford circle anchored at $\tilde\tau = \frac{a}{c}$, while $s(a,c)$ is the Dedekind sum, as was already explained in the introduction.
The formula can also be modified to directly compute also the imaginary part by shifting $s(a,c) \to s(a,c)+\frac{1}{4}$. 

The Rademacher formula \eqref{eq:general Rademacher formula} involves an infinite sum over $c$ and in general the validity of the formula requires that the sum converges absolutely, which imposes a certain boundedness condition on the analytically continued integrand $f(\tau,\tilde{\tau})$. This condition says that for some $\varepsilon>0$
\be 
(\tau+\tilde{\tau})^{2+\varepsilon} \eta(\tau)^{2\varepsilon}\eta(\tilde{\tau})^{2\varepsilon} f(\tau,\tilde{\tau}) \label{eq:boundedness condition}
\ee
is bounded in absolute value by a constant when $\tau$ and $\tilde{\tau}$ are taken to lie on Ford circles (with the constant independent of the Ford circle). 

However, it was observed in \cite{Baccianti:2025gll} that the formula \eqref{eq:general Rademacher formula} also holds in many cases when \eqref{eq:boundedness condition} is only bounded for $\varepsilon=0$. In that case, the sum in \eqref{eq:general Rademacher formula} might only be conditionally convergent. This is the case of interest in this paper. Thus we have to take a small leap of faith that our procedure is actually convergent, for which we already provided strong evidence, for example in Figure~\ref{fig:Baikov-comparison}. The convergence of the analogous formula for the open string was discussed at length in \cite[Appendix B]{Eberhardt:2023xck}.

To apply \eqref{eq:general Rademacher formula} to string amplitudes, it will be necessary to complexify the moduli of the integrand \eqref{eq:4-point function graviton scattering type II regularized}. As above, we denote $\tilde{\tau}=-\bar{\tau}$ and parametrize $z_i=x_i+\tau y_i$. The integrand over $\tau$ becomes a holomorphic function of these complex moduli. The integrand takes the form
\begin{align} 
f(\tau,\tilde{\tau})&= (\Im \tau)^{-5} \int \prod_{i=1}^3 \d^2 z_i\ \prod_{1 \le j<i \le 4}\ | \vartheta_1(z_{ij},\tau)|^{-2s_{ij}}\,  \mathrm{e}^{\frac{2\pi s_{ij} (\Im z_{ij})^2}{\Im \tau}}\\
&=\left(\frac{2i}{\tau+\tilde{\tau}}\right)^2 \int \prod_{i=1}^3 \d x_i\, \d y_i \prod_{1 \le j<i \le 4} \big(-\vartheta_1(x_{ij}+\tau y_{ij},\tau)\vartheta_1(-x_{ij}+\tilde{\tau} y_{ij},\tilde{\tau})\big)^{-s_{ij}}\nonumber\\
&\qquad\times \mathrm{e}^{-\pi i s_{ij} y_{ij}^2(\tau+\tilde{\tau})}
\end{align}
Here we used that $\overline{\vartheta_1(z,\tau)}=-\vartheta_1(-\bar{z},-\bar{\tau})$. In writing this expression one has to be careful since the exponent $s_{ij}$ is potentially non-integer, which introduces branch cuts. The branch is always defined by starting from the real slice where $x_i \in \RR$, $y_i \in \RR$ and $\tilde{\tau}=-\bar{\tau}$ and then employing analytic continuation along the contour of integration. A more practically useful way of keeping track of the contour is to insert the following redundant factor:
\begin{align}
    f(\tau,\tilde{\tau})
&=\left(\frac{2i}{\tau+\tilde{\tau}}\right)^2 \int \prod_{i=1}^3 \d x_i\, \d y_i \prod_{1 \le j<i \le 4} \bigg(-\frac{\vartheta_1(x_{ij}+\tau y_{ij},\tau)\vartheta_1(-x_{ij}+\tilde{\tau} y_{ij},\tilde{\tau})}{\vartheta_1'(0,\tau) \vartheta_1'(0,\tilde{\tau})}\bigg)^{-s_{ij}}\nonumber\\
&\qquad\times \mathrm{e}^{-\pi i s_{ij} y_{ij}^2(\tau+\tilde{\tau})}\ , \label{eq:integrand thetaprime inserted}
\end{align}
which cancels out by momentum conservation. We observe now that for small $x_{ij}$ and $y_{ij}$, the expression in the parenthesis can be Taylor expanded to first order and becomes $(x_{ij}+\tau y_{ij})(x_{ij}-\tilde{\tau} y_{ij})$, which for real $x_i$ and $y_i$ and $\tau,\tilde{\tau} \in \mathbb{H}$ never takes values on the negative real axis, i.e.
\be 
|\arg\big((x_{ij}+\tau y_{ij})(x_{ij}-\tilde{\tau} y_{ij})\big)| <\pi\ .
\ee
Thus, we can choose for any $\tau$ and $\tilde{\tau}$ the principal branch of $(\, {\bullet}\, )^{-s_{ij}}$ in the vicinity of $x_i, y_i \sim 0$ and follow it continuously from there.

The Rademacher formula \eqref{eq:general Rademacher formula} tells us now that 
\be 
\Re A=\sum_{c=1}^\infty \sum_{\begin{subarray}{c} a=0 \\ (a,c)=1 \end{subarray}}^{c-1} A_{a/c}\ , \label{eq:A sum over a and c}
\ee
where
\begin{align}
    A_{a/c}= \int_{ \longrightarrow}\!\!\!\d \tau\int_{C_{a/c}}\!\!\!\d \tilde{\tau} \,  \bigg[\frac{1}{12i}\Big(\tau-\tilde{\tau}+\frac{2a}{c}\Big)+i s(a,c)\bigg]\, f(\tau,\tilde{\tau})
\end{align}
with $f(\tau,\tilde{\tau})$ given in \eqref{eq:integrand thetaprime inserted}. This will be complicated to evaluate, but much easier than the original integral \eqref{eq:4-point function graviton scattering type II}, since we can make the integral over $C_{a/c}$ and $\longrightarrow$ arbitrarily small and only pick up divergent terms near the cusp of moduli space.

\subsection{Modular transformations}
To continue, it is useful to apply a modular transformation to the Ford circle $C_{a/c}$ and map it back to a horizontal contour. Thus, we replace $\tilde{\tau} \to \frac{a \tilde{\tau}+b}{c \tilde{\tau}+d}$ and use the modular properties of the theta-functions. We also reverse the orientation of the contour and accounting for the correct overall sign leads to
\begin{align}
    A_{a/c}&=-\frac{i}{3} \int_{\longrightarrow} \d \tau \int_{\longrightarrow} \frac{\d \tilde{\tau}}{(c \tilde{\tau}+d)^2} \ \left(\tau+\frac{a \tilde{\tau}+b}{c \tilde{\tau}+d}\right)^{-2}\Big(\tau-\frac{a \tilde{\tau}+b}{c \tilde{\tau}+d}+\frac{2a}{c}-12 s(a,c)\Big)\nonumber\\
&\qquad\times \int \prod_{i=1}^3 \d x_i\, \d y_i\!\! \prod_{1 \le j<i \le 4} \!\bigg(-\frac{\vartheta_1(x_{ij}+\tau y_{ij},\tau)\vartheta_1(\frac{(a \tilde{\tau}+b)y_{ij}-(c \tilde{\tau}+d)x_{ij}}{c \tilde{\tau}+d} ,\frac{a\tilde{\tau}+b}{c \tilde{\tau}+d})}{\vartheta_1'(0,\tau) \vartheta_1'(0,\frac{a \tilde{\tau}+b}{c \tilde{\tau}+d})}\bigg)^{\!  -s_{ij}}\nonumber\\
&\qquad\times \mathrm{e}^{-\pi i s_{ij} y_{ij}^2(\tau+\frac{a\tilde{\tau}+b}{c \tilde{\tau}+d})}\\
&=-\frac{i}{3} \int_{\longrightarrow} \d \tau \int_{\longrightarrow} \d \tilde{\tau} \ ((c \tilde{\tau}+d)\tau+a \tilde{\tau}+b)^{-2}\Big(\tau-\frac{a \tilde{\tau}+b}{c \tilde{\tau}+d}\tau+\frac{2a}{c}-12 s(a,c)\Big)\nonumber\\
&\qquad\times \int \prod_{i=1}^3 \d x_i\, \d y_i \!\!\prod_{1 \le j<i \le 4}\! \bigg( \frac{\vartheta_1(x_{ij}+\tau y_{ij},\tau)\vartheta_1((a \tilde{\tau}+b)y_{ij}-(c \tilde{\tau}+d)x_{ij} ,\tilde{\tau})}{-(c \tilde{\tau}+d)\vartheta_1'(0,\tau)\vartheta_1'(0,\tilde{\tau})}\bigg)^{\!  -s_{ij}}\nonumber\\
&\qquad\times \mathrm{e}^{-\pi i s_{ij} y_{ij}^2(\tau+\frac{a\tilde{\tau}+b}{c \tilde{\tau}+d})-\frac{\pi i c s_{ij}}{c \tilde{\tau}+d}((a \tilde{\tau}+b)y_{ij}-(c \tilde{\tau}+d)x_{ij})^2}\\
&=-\frac{i}{3c^2} \int_{\longrightarrow} \d \tau \int_{\longrightarrow} \d \tilde{\tau} \ \Big( \tilde{\tau}\tau-\frac{1}{c^2}\Big)^{-2}\Big(\tau+\frac{1}{c^2 \tilde{\tau}}-12s(a,c)\Big)\int \prod_{i=1}^3 \d x_i\, \d y_i\nonumber\\
&\qquad\times \prod_{1 \le j<i \le 4} \!\bigg(\frac{\vartheta_1(x_{ij}+\tau y_{ij},\tau-\tfrac{a}{c})\vartheta_1(c x_{ij} \tilde{\tau}+\frac{y_{ij}}{c} ,\tilde{\tau}-\tfrac{d}{c})}{\tilde{\tau}\,  \vartheta_1'(0,\tau-\frac{a}{c})\vartheta_1'(0,\tilde{\tau}-\frac{d}{c})}\bigg)^{\! -s_{ij}}\nonumber\\
&\qquad\times \mathrm{e}^{-\pi i s_{ij} (y_{ij}^2 \tau+c^2 x_{ij}^2 \tilde{\tau}+2x_{ij}y_{ij})}\ . \label{eq:Ford circle before q expansion}
\end{align}
In the last line, we shifted $\tilde{\tau} \to \tilde{\tau}-\frac{d}{c}$, $\tau \to \tau-\frac{a}{c}$ and $x_i \to x_i+\frac{a}{c} y_i$. We also used the symmetry properties of the theta function. All the manipulations preserved the reality properties.

The branch is still determined by taking the principal branch for small $x_i$ and $y_i$ and then following it continuously, but it is actually convenient to change this prescription at this point. Let us follow the branch from $x_i \sim y_i \sim 0$ to small, but positive $x_{ij}$ and $y_{ij}$. We also consider large and purely imaginary $\tau$ and $\tilde{\tau}$. In this limit, we have
\be 
\frac{\vartheta_1(x_{ij}+\tau y_{ij},\tau-\tfrac{a}{c})}{  \vartheta_1'(0,\tau-\frac{a}{c})}\sim \sin(\pi(x_{ij}+\tau y_{ij})) \sim \frac{i}{2}\, \mathrm{e}^{-\pi i x_{ij}-\pi i \tau y_{ij}}\ ,
\ee
and similarly for the other factor.
Here we used that one of the terms in the definition of the sine will overwhelm the other one for large $\Im \tau$. This means that the branch of
\be 
\bigg(\frac{\vartheta_1(x_{ij}+\tau y_{ij},\tau-\tfrac{a}{c})}{  i\, \vartheta_1'(0,\tau-\frac{a}{c})}\bigg)^{-s}
\ee
is given by the principal branch in this regime. Therefore, the only effect that this change of prescription has is to multiply \eqref{eq:Ford circle before q expansion} by $1=\prod_{1 \le j<i \le 4}(-\tilde{\tau})^{-s_{ij}}$, which replaces the $\tilde{\tau}$ in the denominator of \eqref{eq:Ford circle before q expansion} by a minus sign.

\subsection{Windings}
The Rademacher formula \eqref{eq:general Rademacher formula} expresses the modular integral as a sum of contributions coming from the Lorentzian singularities of the complexified moduli space $\mathcal{M}_{1,1}^\CC$. In our case, we get further sums since the answer actually splits into contributions from the Lorentzian singularities of $\mathcal{M}_{1,4}^\CC$, which besides the rational number $\frac{a}{c}$ also include certain windings of punctures around one another. We will now see where they come from.

To continue with \eqref{eq:Ford circle before q expansion}, we notice that the integral over $x_i$ and $y_i$ still runs over the interval $[0,1]$ in both variables. The argument of the first theta-function, $x_i+\tau y_i$ runs exactly once over the torus in this integral. However, the imaginary part of the argument of the second theta function, $c x_i \tilde{\tau}+\frac{y_i}{c}$, goes $c$ times around the torus. We will get one polar term for every time it winds around the torus. For this reason, it will be useful to replace
\be 
x_i\to\frac{n_i+x_i}{c}
\ee
with an integer $n_i \in \{0,1,\dots,c-1\}$ and $x_i \in [0,1]$, so that $\frac{n_i+x_i}{c}$ runs over the full interval once. We keep denoting the variable by $x_i$ in order to exhibit the symmetry between $x_i$ and $y_i$. Thus we can write
\begin{align} 
A_{a/c}&=-\frac{i}{3c^5} \!\!\sum_{n_1, n_2, n_3=0}^{c-1}\int_{\longrightarrow} \!\!\d \tau \int_{\longrightarrow}\!\! \d \tilde{\tau} \ \Big( \tilde{\tau}\tau-\frac{1}{c^2}\Big)^{-2}\Big(\tau+\frac{1}{c^2 \tilde{\tau}}-12s(a,c)\Big)\int \prod_{i=1}^3 \d x_i\, \d y_i\nonumber\\
&\qquad\times \prod_{1 \le j<i \le 4} \bigg(-\frac{\vartheta_1(\frac{n_{ij}+x_{ij}}{c}+\tau y_{ij},\tau-\tfrac{a}{c})\vartheta_1((n_{ij}+x_{ij}) \tilde{\tau}+\frac{y_{ij}}{c} ,\tilde{\tau}-\tfrac{d}{c})}{\vartheta_1'(0,\tau-\frac{a}{c})\vartheta_1'(0,\tilde{\tau}-\frac{d}{c})}\bigg)^{-s_{ij}}\nonumber\\
&\qquad\times \mathrm{e}^{-\pi i s_{ij} (y_{ij}^2 \tau+(n_{ij}+x_{ij})^2 \tilde{\tau}+\frac{2}{c}(n_{ij}+x_{ij})y_{ij})}\ . \label{eq:Ford circle before q expansion partitioned}
\end{align}
Since the integral over the torus is translational symmetric, there is an inherited gauge freedom $(n_1,n_2,n_3,n_4) \mapsto (n_1+n,n_2+n,n_3+n,n_4+n)$, which we could use to put $n_4$ to some fixed value.

The next step involves tracking the branch of $(\, \bullet\, )^{-s_{ij}}$ from the region where $x_{ij}$ is slightly bigger than $-n_{ij}$, where we defined the branch to the region close to the region where $x_{ij}$ is slightly bigger than $0$.
For the first theta-function appearing in \eqref{eq:Ford circle before q expansion partitioned} this is simple and the choice of branch becomes manifest by writing
\be 
\bigg(\frac{\vartheta_1(\frac{n+x}{c}+\tau y,\tau-\tfrac{a}{c})}{i\, \vartheta_1'(0,\tau-\frac{a}{c})}\bigg)^{-s}=\mathrm{e}^{\frac{\pi i n s}{c}}\bigg(\frac{\mathrm{e}^{\frac{\pi i n}{c}}\vartheta_1(\frac{n+x}{c}+\tau y,\tau-\tfrac{a}{c})}{i\, \vartheta_1'(0,\tau-\frac{a}{c})}\bigg)^{-s}\ , \label{eq:first branch continuation}
\ee
and we choose the principal branch for small $x>0$ and $y>0$.
For the second theta-function in \eqref{eq:Ford circle before q expansion partitioned}, we use the quasiperiodicity of the theta function, which gives for $n\ge 0$,
\be 
\bigg(\frac{\vartheta_1((n+x) \tilde{\tau} ,\tilde{\tau}-\tfrac{d}{c})}{i\, \vartheta_1'(0,\tilde{\tau}-\frac{d}{c})}\bigg)^{-s}\! =\mathrm{e}^{\pi i s \tau n(n+2 x)+2\pi i s \sum_{m=1}^n \st{\frac{md}{c}}} \bigg(\frac{\mathrm{e}^{\frac{\pi i n d}{c}}\vartheta_1(x \tilde{\tau}+\frac{nd}{c} ,\tilde{\tau}-\tfrac{d}{c})}{i\, \vartheta_1'(0,\tilde{\tau}-\frac{d}{c})}\bigg)^{-s}\ . \label{eq:second branch continuation}
\ee
Here, $\st{x} := x-\lfloor x \rfloor -\frac{1}{2}$ is the sawtooth function.
For $n<0$, the sawtooth sum has to be interpreted as follows,
\begin{align}
    \sum_{m=1}^n \bigst{\frac{md}{c}}\overset{n<0}{:=}\frac{1}{2}+\sum_{m=1}^{-n-1} \bigst{\frac{md}{c}} \ ,\label{eq:negative sawtooth sum}
\end{align}
and we continue to write the sawtooth sum with this convention.
This type of equation was already discussed in \cite{Eberhardt:2023xck} and thus we will not repeat the derivation, but only explain the idea behind it. It can be proven by following the argument of the theta-function for large $\tilde{\tau}$ continuously in $x$. The sawooth function enters via the following mechanism. By following the branch of $(1-\mathrm{e}^{x+i \varphi})^{-s}$, smoothly from $x<0$ to $x>0$, the argument changes rapidly around $x \sim 0$. For $x<0$, we simply pick the principal branch. For $x>0$, the correct answer is extracted by writing
\be 
(1-\mathrm{e}^{x+2\pi i \varphi})^{-s}=\mathrm{e}^{-2\pi i s \st{\varphi}-s x} (1-\mathrm{e}^{-x-i \varphi})^{-s}\ .
\ee
The branch jumps occurs precisely when $\varphi \in \ZZ$, since the branch point lies on the real axis in that case.

If we use \eqref{eq:first branch continuation} and \eqref{eq:second branch continuation}, we find that we can write
\begin{align}
    A_{a/c}&=-\frac{i}{3c^5} \!\!\sum_{n_1,n_2,n_3=0}^{c-1}\!\! \mathrm{e}^{2\pi i \sum_{1 \le j<i\le 4} s_{ij} \big[\frac{n_{ij}}{2c}+\sum_{m=1}^{n_{ij}} \st{\frac{md}{c}}\big]} \int_{\longrightarrow} \!\!\d \tau \int_{\longrightarrow} \!\!\d \tilde{\tau} \ \Big( \tilde{\tau}\tau-\frac{1}{c^2}\Big)^{-2}\nonumber\\
    &\qquad\times\Big(\tau+\frac{1}{c^2 \tilde{\tau}}-12s(a,c)\Big)\int \prod_{i=1}^3 \d x_i\, \d y_i \prod_{1 \le j<i \le 4} \mathrm{e}^{-\pi i s_{ij} (y_{ij}^2 \tau+x_{ij}^2 \tilde{\tau}+\frac{2x_{ij}y_{ij}}{c})}\nonumber\\
    &\qquad\times \bigg(-\frac{\mathrm{e}^{\frac{\pi i n_{ij}(1+d)}{c}}\vartheta_1(y_{ij}\tau+\frac{n_{ij}+x_{ij}}{c},\tau-\tfrac{a}{c})\vartheta_1(x_{ij} \tilde{\tau}+\frac{n_{ij}d+y_{ij}}{c} ,\tilde{\tau}-\tfrac{d}{c})}{  \vartheta_1'(0,\tilde{\tau}-\frac{a}{c})\vartheta_1'(0,\tilde{\tau}-\frac{d}{c})}\bigg)^{-s_{ij}}
\end{align}
with the branch choice as discussed above. We see that this expression is remarkably symmetric under $y_i \leftrightarrow x_i$, $\tau \leftrightarrow \tilde{\tau}$, $a \leftrightarrow d$ and $n_i \leftrightarrow n_i a$. We will see this symmetry more prominently in the following. 

\subsection{Change of variables}
To continue, it is useful to observe that the arguments of the theta-functions only depend on the variables
\be 
z_{ij}\tau =y_{ij}\tau+\frac{x_{ij}}{c}\ , \qquad \tilde{z}_{ij} \tilde{\tau}=x_{ij} \tilde{\tau}+\frac{y_{ij}}{c}\ . \label{eq:z ztilde definition}
\ee
It is thus useful to change variables accordingly. The contour will of course not factorize directly and we will write $\int_{\mathbb{T}^2_c} \d z_i \, \d \tilde{z}_i$ for the integral over the appropriate contour. Including the appropriate Jacobian, we thus find
\begin{align}
    A_{a/c}&=-\frac{i}{3c^5} \!\!\sum_{n_1,n_2,n_3=0}^{c-1}\!\! \mathrm{e}^{2\pi i \sum_{1 \le j<i\le 4} s_{ij} \big[\frac{n_{ij}}{2c}+\sum_{m=1}^{n_{ij}} \st{\frac{md}{c}}\big]}  \int_{\longrightarrow} \frac{\d \tau}{\tau^2} \int_{\longrightarrow} \frac{\d \tilde{\tau}}{\tilde{\tau}^2} \ \Big(1-\frac{1}{c^2 \tau \tilde{\tau}}\Big)^{-5}\nonumber\\
    &\qquad\times\Big(\tau+\frac{1}{c^2 \tilde{\tau}}-12s(a,c)\Big)  \int_{\mathbb{T}^2_c} \prod_{i=1}^3\d z_i \, \d \tilde{z}_i\prod_{1 \le j<i \le 4} \mathrm{e}^{-\frac{\pi i s_{ij}(z_{ij}^2 \tau+ \tilde{z}_{ij}^2 \tilde{\tau}-2c^{-1} z_{ij} \tilde{z}_{ij})}{1-\frac{1}{c^2 \tau \tilde{\tau}}}}\nonumber\\
    &\qquad\times\bigg(-\frac{\mathrm{e}^{\frac{\pi i n_{ij}(1+d)}{c}}\vartheta_1(z_{ij}\tau +\frac{n_{ij}}{c},\tau-\tfrac{a}{c})\vartheta_1(\tilde{z}_{ij}\tilde{\tau}+\frac{n_{ij}d}{c} ,\tilde{\tau}-\tfrac{d}{c})}{  \vartheta_1'(0,\tilde{\tau}-\frac{a}{c})  \vartheta_1'(0,\tilde{\tau}-\frac{d}{c})}\bigg)^{-s_{ij}}\ . \label{eq:Ford circle winding contribution}
\end{align}

\subsection{Tropicalization and \texorpdfstring{$q$}{q}-expansion}
To proceed, we notice that only a finite number of terms in the $q=\mathrm{e}^{2\pi i \tau}$ and $\tilde{q}=\mathrm{e}^{2\pi i \tilde{\tau}}$ expansion of the integrand will contribute to the to the integral. These are those which can have negative exponent for some choice of the other moduli, since those with positive exponent will die off near the cusp. By pushing the imaginary part of the horizontal contour over $\tau$ and $\tilde{\tau}$ to infinity, we can make their contribution arbitrarily small and thus they don't contribute. In analogy to the terminology used for holomorphic modular objects, we call these the polar terms.
To understand the structure of these polar terms, it is useful to pick out one term in the $q$-expansion and analyze it further. For this, we may ignore all phases and only keep track of the exponent of $q$. We see that the structure of the polar terms completely factorizes in $q$ and $\tilde{q}$ and each chiral half is as for the open string \cite{Eberhardt:2023xck}. Of course, the $z_i$-contour doesn't quite run over real $z_i$'s. But notice that only the imaginary part of $z_{ij} \tau$ controls the growth when $\Im \tau \to \infty$ and $\Im (z_{ij} \tau)=y_{ij} \Im \tau$. Thus, we can write the condition on the growth as if $z_i$ was real.

Each term in the $q$-expansion has the exponent \cite[Equation~(5.12)]{Eberhardt:2023xck}
\begin{multline} 
\Trop_{m_\L,m_\D,m_\R,m_\U}=-\frac{1}{2}\sum_{i>j} s_{ij} z_{ij}(z_{ij}-1)+
\left(\begin{cases}
m_\L z_{21} &\;\text{if}\quad z_{21}>0  \\
(m_\L-s) z_{21} &\;\text{if}\quad z_{21}<0
\end{cases}\right) + m_\D z_{32}\\
+\left(\begin{cases}
m_\R z_{43} &\;\text{if}\quad z_{43}>0  \\
(m_\R-s)z_{43} &\;\text{if}\quad z_{43}<0
\end{cases}\right) +m_\U(1-z_{41}) \label{eq:Trop}
\end{multline}
for some non-negative integers $m_\L$, $m_\D$, $m_\R$ and $m_\U$, which have the interpretation of masses squared, $m_i = \m^2_i$ of intermediate string states. The condition $\Trop_{m_\L,m_\D,m_\R,m_\U}<0$ defines a a certain region in the $z_{i}$ parameter space. For $s$-channel kinematics, this region contains the hypersurface $z_{21}=0$ and $z_{43}=0$, but does \emph{not} contain the other hypersurfaces $z_{ij}=0$. This means that as for the open string, one can $q$-expand also the zero modes of the theta-functions that do not involve $z_{21}$ and $z_{43}$.

The $q$-expansion completely factorizes. From the infinite product representation of the theta-functions, we can write
\begin{multline}
    \prod_{1 \le j<i \le 4} \bigg(\frac{\mathrm{e}^{\frac{\pi i n_{ij}}{c}}\vartheta_1(z_{ij}\tau+\frac{n_{ij}}{c},\tau-\tfrac{a}{c})}{i\, \vartheta_1'(0,\tau-\frac{a}{c})}\bigg)^{-s_{ij}} =\prod_{1 \le j<i \le 4} q^{\frac{1}{2}s_{ij}z_{ij}} \\ \times\prod_{\ell=0}^\infty \big(1-q^{\ell+z_{ij}} \mathrm{e}^{\frac{2\pi i (n_{ij}-a \ell)}{c}}\big)^{-s_{ij}}\prod_{\ell=1}^\infty\big(1-q^{\ell-z_{ij}} \mathrm{e}^{-\frac{2\pi i (n_{ij}+a \ell)}{c}}\big)^{-s_{ij}}\ .
\end{multline}
We then $q$-expand everything except for the terms with $ij=21, 43$ and $\ell=0$. This gives 
\begin{align}
    &\!\!\!\!\!\prod_{1 \le j<i \le 4} \bigg(\frac{\mathrm{e}^{\frac{\pi i n_{ij}}{c}}\vartheta_1(z_{ij}\tau+\frac{n_{ij}}{c},\tau-\tfrac{a}{c})}{i\, \vartheta_1'(0,\tau-\frac{a}{c})}\bigg)^{-s_{ij}}\nonumber\\
    &= \prod_{1 \le j<i \le 4} q^{\frac{1}{2}s_{ij}z_{ij}}  \sum_{
        m_\L,m_\D,m_\R,m_\U \ge 0} Q_{m_\L,m_\D,m_\R,m_\U}q^{m_\L z_{21}+m_\D z_{32}+m_\R z_{43}+m_\U (1-z_{41})} \nonumber\\
    &\qquad\times \mathrm{e}^{\frac{2\pi i}{c} (n_{21} m_\L+n_{32} m_\D+n_{43} m_\R+n_{14} m_\U)-\frac{2\pi i a m_\U}{c}} \big(1-\mathrm{e}^{\frac{2\pi i}{c} n_{21}} q^{z_{21}}\big)^{-s}\big(1-\mathrm{e}^{\frac{2\pi i}{c}n_{43}} q^{z_{43}}\big)^{-s}\ .
\end{align}
The exponent of $q$ when taking also the Gaussian term in \eqref{eq:Ford circle winding contribution} into account gives the exponent \eqref{eq:Trop}.
Here, $Q_{m_\L,m_\D,m_\R,m_\U}$ coincides with the same polynomials in $s$ and $t$ encountered for the open string \cite{Eberhardt:2023xck}, where we defined them via their generating function
\begin{align}  
Q_{m_\L,m_\D,m_\R,m_\U}(s,t) &=[q_\L^{m_\L} q_\D^{m_\D}q_\R^{m_\R}q_\U^{m_\U}] \prod_{\ell=1}^\infty \prod_{a=\L,\R}(1-q^\ell q_a^{-1})^{-s}(1-q^\ell q_a)^{-s}\nonumber\\
&\qquad\times\prod_{a=\D,\U}(1-q^\ell q_a^{-1})^{-t}(1-q^{\ell-1} q_a)^{-t}\nonumber\\
&\qquad\times \prod_{a=\L,\R}(1-q^{\ell}q_a^{-1} q_\D^{-1})^{-u} (1-q^{\ell-1}q_a q_\D)^{-u}\ . \label{eq:QmL,mD,mR,mU definition}
\end{align}
We can then insert this expansion into \eqref{eq:Ford circle winding contribution} for both theta-function factors and write the result as a finite sum over the integers $m_a$ and $\tilde{m}_a$. Let us factor out some of the phase prefactors and write
\begin{align}
    A_{a/c}&= -\frac{i}{3c^5} \!\!\sum_{n_1,n_2,n_3=0}^{c-1}\!\! \mathrm{e}^{2\pi i \sum_{1 \le j<i\le 4} s_{ij} \big[\frac{n_{ij}}{2c}+\sum_{m=1}^{n_{ij}} \st{\frac{md}{c}}\big]} \nonumber\\
    &\qquad\times \sum_{\begin{subarray}{c}
        m_\L,m_\D,m_\R,m_\U \ge 0 \\
\tilde{m}_\L,\tilde{m}_\D,\tilde{m}_\R,\tilde{m}_\U \ge 0
    \end{subarray}} Q_{m_\L,m_\D,m_\R,m_\U}Q_{\tilde{m}_\L,\tilde{m}_\D,\tilde{m}_\R,\tilde{m}_\U} \nonumber\\
    &\qquad \times \mathrm{e}^{\frac{2\pi i}{c} (n_{21} m_\L+n_{32} m_\D+n_{43} m_\R+n_{14} m_\U)-\frac{2\pi i a m_\U}{c}} \nonumber\\
    &\qquad\times \mathrm{e}^{\frac{2\pi i d}{c} (n_{21} \tilde{m}_\L+n_{32} \tilde{m}_\D+n_{43} \tilde{m}_\R+n_{14} \tilde{m}_\U)-\frac{2\pi i d \tilde{m}_\U}{c}} A_{a/c}^{\boldsymbol{n}_a,\boldsymbol{m}_a,\boldsymbol{\tilde{m}}_a}\ , \label{eq:sum over all m and mtilde}
\end{align} 
with
\begin{align}
    A_{a/c}^{\boldsymbol{n}_a,\boldsymbol{m}_a,\boldsymbol{\tilde{m}}_a}&= \int_{\longrightarrow} \frac{\d \tau}{\tau^2} \int_{\longrightarrow} \frac{\d \tilde{\tau}}{\tilde{\tau}^2} \ \Big(1-\frac{1}{c^2 \tau \tilde{\tau}}\Big)^{-5}\Big(\tau+\frac{1}{c^2 \tilde{\tau}}-12s(a,c)\Big) \prod_{i=1}^3 \int_{\mathbb{T}^2_c} \d z_i \, \d \tilde{z}_i\nonumber\\
&\qquad\times \prod_{1 \le j<i \le 4} \mathrm{e}^{-\pi i s_{ij}(1-\frac{1}{c^2 \tau \tilde{\tau}})^{-1}(z_{ij}^2 \tau+ \tilde{z}_{ij}^2 \tilde{\tau}-2c^{-1} z_{ij} \tilde{z}_{ij})+\pi i s_{ij} (\tau z_{ij}+\tilde{\tau}\tilde{z}_{ij})}\nonumber\\
&\qquad\times q^{m_\L z_{21}+m_\D z_{32}+m_\R z_{43}+m_\U (1-z_{41})}\tilde{q}^{\tilde{m}_\L \tilde{z}_{21}+\tilde{m}_\D \tilde{z}_{32}+\tilde{m}_\R \tilde{z}_{43}+\tilde{m}_\U (1-\tilde{z}_{41})}\nonumber\\
&\qquad\times \big((1-\mathrm{e}^{\frac{2\pi i}{c} n_{21}} q^{z_{21}})(1-\mathrm{e}^{\frac{2\pi id}{c} n_{21}} \tilde{q}^{\tilde{z}_{21}})\big)^{-s}\nonumber\\
&\qquad\times \big((1-\mathrm{e}^{\frac{2\pi i}{c}n_{43}} q^{z_{43}})(1-\mathrm{e}^{\frac{2\pi id}{c}n_{43}} \tilde{q}^{\tilde{z}_{43}})\big)^{-s}\ .
\end{align}
\subsection{Evaluating the integrals} \label{subsec:evaluating integrals}
To continue, we have to evaluate the involved integrals. We follow a similar strategy as for the open string \cite{Eberhardt:2023xck}, where the trick was to integrate in a certain Gaussian integral in order to make the integral over $z_{21}$ and $z_{43}$, a standard beta-function integral. To achieve this, we first make the change of variables
\begin{align}
    z_\L=z_{21}\ , \qquad z_\R=z_{43}\ , \qquad \tilde{z}_\L=\tilde{z}_{21}\ , \qquad \tilde{z}_\R=\tilde{z}_{43}\ .
\end{align}
For the remaining $z$-coordinate, we change variables to $t_\L$ and $\tilde{t}_\L$, related by
\begin{align}
    z_{31}&=\frac{s(1+z_\L-z_\R)+m_\U-m_\D}{2s}-\frac{(\tilde{m}_\D+\tilde{m}_\U-s-2\tilde{t}_\L+2t \tilde{z}_\R)+s(\tilde{z}_\L+\tilde{z}_\R)}{2s}\frac{\sqrt{\tilde{\tau}}}{\sqrt{\tau}} \nonumber\\
    &\qquad+\frac{s+2t_\L-m_\D-m_\U}{2s c \sqrt{\tau}\sqrt{\tilde{\tau}}}+\frac{s+\tilde{m}_\U-\tilde{m}_\D}{2 s c \tau}\ , \\
    \tilde{z}_{31}&=\frac{s(1+\tilde{z}_\L-\tilde{z}_\R)+\tilde{m}_\U-\tilde{m}_\D}{2s}-\frac{(m_\D+m_\U-s-2t_\L+2t z_\R)+s(z_\L+z_\R)}{2s}\frac{\sqrt{\tau}}{\sqrt{\tilde{\tau}}} \nonumber\\
    &\qquad+\frac{s+2\tilde{t}_\L-\tilde{m}_\D-\tilde{m}_\U}{2s c \sqrt{\tau}\sqrt{\tilde{\tau}}}+\frac{s+m_\U-m_\D}{2 s c \tilde{\tau}}\ .
\end{align}
We also integrate in the Gaussian integral
\begin{align}
    1&=\frac{s \sqrt{\frac{1}{c^2}-\tau \tilde{\tau}}}{2 t u}\int \d t_\R\, \d \tilde{t}_\R\, \mathrm{e}^{\frac{\pi i \tau}{2 s t u}(s (t_\R+\frac{\tilde{t}_\R}{c \tau})-(s+2t)(t_\L+\frac{\tilde{t}_\L}{c \tau})-2 t u z_\R+t(m_\D+\frac{\tilde{m}_\D}{c \tau}+m_\U+\frac{\tilde{m}_\U}{c \tau})-s t (1+\frac{1}{c \tau}))^2} \nonumber\\
    &\qquad \times \mathrm{e}^{\frac{2\pi i \tilde{\tau}}{4 s tu}(1-\frac{1}{c^2 \tau \tilde{\tau}})^{-1}((s \tilde{t}_\R-(s+2t) \tilde{t}_\L+(\tilde{m}_\D+\tilde{m}_\U-s)t)(1-\frac{1}{c^2 \tau \tilde{\tau}})-2 t u(\tilde{z}_\R-\frac{z_\R}{c \tilde{\tau}}))^2}\ . \label{eq:integrating in 1}
\end{align}
Including the Jacobians, this procedure yields
\begin{align}
    &\prod_{i=1}^3 \int_{\mathbb{T}^2_c} \d z_i \, \d \tilde{z}_i\prod_{1 \le j<i \le 4} \mathrm{e}^{-\pi i s_{ij}(1-\frac{1}{c^2 \tau \tilde{\tau}})^{-1}(z_{ij}^2 \tau+ \tilde{z}_{ij}^2 \tilde{\tau}-2c^{-1} z_{ij} \tilde{z}_{ij})+\pi i s_{ij} (\tau z_{ij}+\tilde{\tau}\tilde{z}_{ij})}\nonumber\\
&\qquad\times q^{m_\L z_{21}+m_\D z_{32}+m_\R z_{43}+m_\U (1-z_{41})}\tilde{q}^{\tilde{m}_\L \tilde{z}_{21}+\tilde{m}_\D \tilde{z}_{32}+\tilde{m}_\R \tilde{z}_{43}+\tilde{m}_\U (1-\tilde{z}_{41})}\nonumber\\
&\qquad\times \big((1-\mathrm{e}^{\frac{2\pi i}{c} n_{21}} q^{z_{21}})(1-\mathrm{e}^{\frac{2\pi id}{c} n_{21}} \tilde{q}^{\tilde{z}_{21}})\big)\big)^{-s}\big((1-\mathrm{e}^{\frac{2\pi i}{c}n_{43}} q^{z_{43}})(1-\mathrm{e}^{\frac{2\pi id}{c}n_{43}} \tilde{q}^{\tilde{z}_{43}})\big)^{-s}\nonumber\\
&\quad=\frac{\sqrt{- \tau \tilde{\tau}}}{2 s tu} \Big(1-\frac{1}{c^2 \tau \tilde{\tau}}\Big)^{\frac{3}{2}} \int \d t_\L\, \d \tilde{t}_\L\, \d t_\R\, \d \tilde{t}_\R\, \d z_\L\, \d \tilde{z}_\L\, \d z_\R\, \d \tilde{z}_\R\nonumber\\
&\qquad\times\mathrm{e}^{-\frac{4\pi i}{c} P_{m_\D,m_\U,\tilde{m}_\D,\tilde{m}_\U}(t_\L,t_\R,\tilde{t}_\L,\tilde{t}_\R)}\nonumber\\
&\qquad\times q^{-z_\L(t_\L-m_\L)-z_\R (t_\R-m_\R)-P_{m_\D,m_\U}(t_\L,t_\R)} \tilde{q}^{-\tilde{z}_\L(t_\L-\tilde{m}_\L)-\tilde{z}_\R(t_\R-\tilde{m}_\R)-P_{\tilde{m}_\D,\tilde{m}_\U}(\tilde{t}_\L,\tilde{t}_\R)} \nonumber\\
&\qquad\times \big((1-\mathrm{e}^{\frac{2\pi i}{c} n_{21}} q^{z_\L})(1-\mathrm{e}^{\frac{2\pi id}{c} n_{21}} \tilde{q}^{\tilde{z}_\L})\big)^{-s}\big((1-\mathrm{e}^{\frac{2\pi i}{c}n_{43}} q^{z_\R})(1-\mathrm{e}^{\frac{2\pi id}{c}n_{43}} \tilde{q}^{\tilde{z}_\R})\big)^{-s}\ , \label{eq:change of variables z integral}
\end{align}
where
\begin{align}
    P_{m_\D,m_\U}(t_\L,t_\R)&=-\frac{1}{4 s tu} \big(s^2 (t_\L - t_\R)^2 + 
   2 s  t  (m_\D + m_\U - s) (t_\L + t_\R) - 4  s  t  t_\L  t_\R \nonumber\\
   &\qquad- 
   4  s  t  m_\D  m_\U + t^2 (m_\D - m_\U)^2 - s  t^2 (2 m_\D + 2 m_\U - s)\big) \label{eq:P polynomial}
\end{align}
is the kinematical kernel given by a ratio of two Gram determinants (see \cite[Equation~(3.26)]{Eberhardt:2023xck}) and
\begin{align}
    &P_{m_\D,m_\U,\tilde{m}_\D,\tilde{m}_\U}(t_\L,t_\R,\tilde{t}_\L,\tilde{t}_\R)\nonumber\\
    &\qquad=-\frac{1}{4  s  t u}\big( (s^2 (t_\L - t_\R) (\tilde{t}_\L - \tilde{t}_\R) + 
   s  t  (\tilde{m}_\D + \tilde{m}_\U - s) (t_\L + t_\R) \nonumber\\
   &\qquad\qquad+ 
   s  t  (m_\D + m_\U - s) (\tilde{t}_\L + \tilde{t}_\R) - 2  s  t  (t_\L  \tilde{t}_\R + \tilde{t}_\L  t_\R) - 
   2 s  t  (m_\D  \tilde{m}_\U + \tilde{m}_\D  m_\U) \nonumber\\
   &\qquad\qquad+ t^2 (m_\D - m_\U) (\tilde{m}_\D - \tilde{m}_\U) + 
   2s^2t(m_\U+\tilde{m}_\U)\nonumber\\
   &\qquad\qquad+st^2(s+m_\U+\tilde{m}_\U-m_\D-\tilde{m}_\D)\big) \label{eq:P2 polynomial}
\end{align}
is a sort of cross kernel, for which we do not have a clear kinematical interpretation.

The main point of this operation is that the $z_\L$, $z_\R$, $\tilde{z}_\L$, $\tilde{z}_\R$, as well as the $\tau$ and $\tilde{\tau}$ behavior is simple enough to do these integrals analytically. We should note that the integration contour in \eqref{eq:change of variables z integral} over the new variables is in principle complicated. However, we can deform the contour freely as long as we don't encounter any singularities. In $t_\L$, $t_\R$, $\tilde{t}_\L$ and $\tilde{t}_\R$, there are no singularities and we may simply extend the integration region over $\RR$ in all four integrals. This requires extending the integral, but the Trop function \eqref{eq:Trop} is positive in the regions where we extend the integral, which means that the difference will not contribute to the $\tau$-integral. Since the tropical structure is the double copy of the open string case, the argument for this is completely analogous to what is discussed in \cite{Eberhardt:2023xck}.

\paragraph{$n_{21} \ne 0$.} For $z_\L$, $\tilde{z}_\L$ (and similarly for $z_\R$, $\tilde{z}_\R$), there are singularities in the integrand. For $n_{21}=0$, these are located at $z_\L=\tilde{z}_\L=0$, while they are far away from the origin for $n_{21} \ne 0$. This means that for $n_{21} \ne 0$, we can simply rotate the contour to the real axis and integrate over $z_\L$ and $\tilde{z}_\L$ independently. In that case, we thus have to compute the integral
\begin{align} 
&\int_{-\infty}^\infty \d z_\L\ q^{-z_\L(t_\L-m_\L)} \big(1-\mathrm{e}^{\frac{2\pi i}{c} n_{21}} q^{z_\L}\big)^{-s}\nonumber\\
&\qquad=\frac{i\, \mathrm{e}^{2\pi i (t_\L-m_\L) \st{\frac{n_{21}}{c}}}}{2\pi \tau} \int_{0}^\infty \d \alpha\ \alpha^{m_\L-t_\L-1}(1+\alpha)^{-s} \\
&\qquad=\frac{i\, \mathrm{e}^{2\pi i (t_\L-m_\L) \st{\frac{n_{21}}{c}}}}{2\pi \tau} \frac{\Gamma(m_\L-t_\L)\Gamma(s+t_\L-m_\L)}{\Gamma(s)}\ .
\end{align}
This integral is also needed in the open string case and its derivation is explained in more detail in \cite{Eberhardt:2023xck}.

\paragraph{$n_{21}=0$.} For $n_{21}=0$, we will need to be somewhat more careful with the contours similar to the derivation of the KLT formula \cite{Kawai:1985xq}. We are interested in the integral
\begin{align}
    I=\int \d z_\L \, \d \tilde{z}_\L \ q^{-z_\L(t_\L-m_\L)}\tilde{q}^{-\tilde{z}_\L (t_\L-\tilde{m}_\L)} \big((1- q^{z_\L})(1- \tilde{q}^{\tilde{z}_\L})\big)^{-s}\ .
\end{align}
To simplify a bit, we notice that we can easily get rid off the $\tau$ and $\tilde{\tau}$ dependence by rescaling $z_\L \to \frac{i z}{2\pi \tau}$ and $\tilde{z}_\L \to \frac{i \tilde{z}}{2\pi \tilde{\tau}}$. 
This brings the integral into the form
\begin{align}
    I=-\frac{1}{4\pi^2 \tau \tilde{\tau}}\int \d z \, \d \tilde{z}\ \mathrm{e}^{z(t_\L-m_\L)+\tilde{z}(\tilde{t}_\L-\tilde{m}_\L)} \big((1-\mathrm{e}^{-z})(1-\mathrm{e}^{-\tilde{z}})\big)^{-s} \label{eq:KLT integral tau rescaled}
\end{align}
The contour is not quite the real axis, since it originates from the change of variables \eqref{eq:z ztilde definition}, where $x_i$ and $y_i$ were real. This change of variables only tells us how to choose the contour and in particular we can choose any value of $\tau$, $\tilde{\tau}$ and $c$ in \eqref{eq:z ztilde definition}. Taking $c$ asymptotically large, this means that we can specify the contour as
\be 
z=y-i \varepsilon x\ , \qquad \tilde{z}=x-i \varepsilon y
\ee
with $\varepsilon>0$ and $x,y \in \RR$. 

We then divide the integral \eqref{eq:KLT integral tau rescaled} into four regions where $\Re z$ and $\Re \tilde{z}$ are negative or positive. In these regions we have to figure out the correct phase. This can be done for $x$ and $y$ near the origin and we can Taylor expand to first order, $1-\mathrm{e}^{-z} \approx z$. Thus for the purpose of figuring out the correct branch we consider
\be 
(z \tilde{z})^{-s}=\big((y-i \varepsilon x)(x-i \varepsilon y)\big)^{-s}=r^{-2s} \big((\sin \varphi-i \varepsilon \cos \varphi)(\cos \varphi-i \varepsilon \sin \varphi)\big)^{-s}\ ,
\ee
where we changed to polar coordinates in the last step. As we vary $\varphi$, the argument of the function is
\begin{align}
    \arg \big((\sin \varphi-i \varepsilon \cos \varphi)(\cos \varphi-i \varepsilon \sin \varphi)\big)=\begin{cases}
        0 \ , & \sin(2\varphi
        )>0\ ,  \\
        -\pi \ , & \sin(2\varphi
        )<0\ .
    \end{cases}
\end{align}
This means that the correct branch for $\big((1-\mathrm{e}^{-z})(1-\mathrm{e}^{-\tilde{z}})\big)^{-s}$ is the principal branch for $z \tilde{z}>0$ and we can rotate the contour on the real axis without danger there, while for $z \tilde{z}<0$, we define
\be 
\big((1-\mathrm{e}^{-z})(1-\mathrm{e}^{-\tilde{z}})\big)^{-s}=\mathrm{e}^{\pi i s}\big(-(1-\mathrm{e}^{-z})(1-\mathrm{e}^{-\tilde{z}})\big)^{-s} .
\ee
Changing variables once more, we have
\begin{align}
    I&=-\frac{1}{4\pi^2 \tau \tilde{\tau}} \int_0^\infty \d \alpha\int_0^\infty \d \tilde{\alpha}\, \alpha^{m_\L-t_\L-1} \tilde{\alpha}^{\tilde{m}_\L-\tilde{t}_\L-1} \mathrm{e}^{\pi i s \delta_{(\alpha-1)(\tilde{\alpha}-1)<0}}\big|(1-\alpha)(1-\tilde{\alpha})\big|^{-s} \\
    &=-\frac{(-1)^{m_\L+\tilde{m}_\L}(\mathrm{e}^{\pi i (t_\L-\tilde{t}_\L)}+\mathrm{e}^{\pi i (\tilde{t}_\L-t_\L)}-\mathrm{e}^{\pi i (t_\L+\tilde{t}_\L)}-\mathrm{e}^{-\pi i (2s+t_\L+\tilde{t}_\L)})}{4\pi^2 \tau \tilde{\tau} (1-\mathrm{e}^{-2\pi i s})} \nonumber\\
    &\qquad \times \frac{\Gamma(m_\L-t_\L)\Gamma(s+t_\L-m_\L)\Gamma(\tilde{m}_\L-\tilde{t}_\L)\Gamma(s+\tilde{t}_\L-\tilde{m}_\L)}{\Gamma(s)^2}\ .
\end{align}
\paragraph{$\tau$ and $\tilde{\tau}$-integral.} After performing the integrals over the $z$'s, the $\tau$ and $\tilde{\tau}$ dependence also becomes sufficiently simple to perform explicitly. Let us collect all factors that are still $\tau$ and $\tilde{\tau}$ dependent. They are
\begin{align}
    I_2&=\int_{\longrightarrow} \frac{\d \tau}{(-i \tau)^{\frac{7}{2}}}\int_{\longrightarrow} \frac{\d \tilde{\tau}}{(-i \tau)^{\frac{7}{2}}}\,  \Big(1-\frac{1}{c^2 \tau \tilde{\tau}}\Big)^{-\frac{7}{2}}\Big(\tau+\frac{1}{c^2 \tilde{\tau}}-12s(a,c)\Big) q^{-P} \tilde{q}^{-\tilde{P}}\\
    &=\int_{\longrightarrow} \d \tau \int_{\longrightarrow} \d \tilde{\tau}\  \Big(\frac{1}{c^2}-\tau \tilde{\tau}\Big)^{-\frac{7}{2}} \Big(\tau+\frac{1}{c^2 \tilde{\tau}}-12 s(a,c)\Big)q^{-P} \tilde{q}^{-\tilde{P}}\ .
\end{align}
where we wrote $P=P_{m_\D,m_\U}$ and $\tilde{P}=P_{\tilde{m}_\D,\tilde{m}_\U}$. This integral is of the form worked out in \cite{Baccianti:2025gll} and can be evaluated in terms of Bessel function. In the notation of \cite{Baccianti:2025gll},
\begin{align}
    I_2&=c^{\frac{7}{2}} \big(\mathcal{I}^{(2)}_c(\tfrac{7}{2},2\pi P, 2\pi \tilde{P})-12s(a,c)\mathcal{I}^{(1)}_c(\tfrac{7}{2},2\pi P, 2\pi \tilde{P})\big) \\
    &=\frac{2^{\frac{15}{2}}\pi^4 c^{\frac{3}{2}}(P \tilde{P})^{\frac{5}{4}}}{15}\Big[i \sqrt{\tfrac{\tilde{P}}{P}}\Big(J_{\frac{3}{2}}\big(\tfrac{4\pi \sqrt{P \tilde{P}}}{c}\big)-J_{\frac{7}{2}}\big(\tfrac{4\pi \sqrt{P \tilde{P}}}{c}\big)\Big)-12c \, s(a,c) J_{\frac{5}{2}}\big(\tfrac{4\pi \sqrt{P \tilde{P}}}{c}\big)\Big]\ .
\end{align}
It vanishes for $P<0$ or $\tilde{P}<0$. This reflects the fact that only terms with $P<0$ and $\tilde{P}<0$ are polar terms and contribute. Thus the integration region in the variables $(t_\L,t_\R,\tilde{t}_\L,\tilde{t}_\R)$ gets localized to these ellipses.
\paragraph{Summary.} If we combine these explicit evaluations, we obtain
\begin{align}
    &A_{a/c}^{\boldsymbol{n}_a,\boldsymbol{m}_a,\tilde{\boldsymbol{m}}_a}\nonumber\\
    &\quad=\frac{2^{\frac{5}{2}} c^{\frac{3}{2}}}{15s t u}\int_{P>0} \d t_\L \, \d t_\R\, \int_{\tilde{P}>0} \d \tilde{t}_\L\, \d \tilde{t}_\R \, \mathrm{e}^{-\frac{4\pi i}{c} P^{(2)}} (P \tilde{P})^{\frac{5}{4}}\nonumber\\
    &\qquad\times\Big[i \sqrt{\tfrac{\tilde{P}}{P}}\Big(J_{\frac{3}{2}}\big(\tfrac{4\pi \sqrt{P \tilde{P}}}{c}\big)-J_{\frac{7}{2}}\big(\tfrac{4\pi \sqrt{P \tilde{P}}}{c}\big)\Big)-12c \, s(a,c) J_{\frac{5}{2}}\big(\tfrac{4\pi \sqrt{P \tilde{P}}}{c}\big)\Big] \nonumber\\
    &\qquad\times \frac{\Gamma(m_\L-t_\L)\Gamma(s+t_\L-m_\L)\Gamma(m_\R-t_\R)\Gamma(s+t_\R-m_\R)}{\Gamma(s)^2} \nonumber \\
    &\qquad\times \frac{\Gamma(\tilde{m}_\L-\tilde{t}_\L)\Gamma(s+\tilde{t}_\L-\tilde{m}_\L)\Gamma(\tilde{m}_\R-\tilde{t}_\R)\Gamma(s+\tilde{t}_\R-\tilde{m}_\R)}{\Gamma(s)^2} \nonumber \\
    &\qquad\times \begin{cases}
        \mathrm{e}^{2\pi i (t_\L-m_\L)\st{\frac{n_{21}}{c}}+2\pi i (\tilde{t}_\L-\tilde{m}_\L) \st{\frac{d n_{21}}{c}}}\ , & n_{21} \ne 0 \\
        \frac{(-1)^{m_\L+\tilde{m}_\L}(\mathrm{e}^{\pi i (t_\L-\tilde{t}_\L)}+\mathrm{e}^{\pi i (\tilde{t}_\L-t_\L)}-\mathrm{e}^{\pi i (t_\L+\tilde{t}_\L)}-\mathrm{e}^{-\pi i (2s+t_\L+\tilde{t}_\L)})}{ 1-\mathrm{e}^{-2\pi i s}}\ , & n_{21}=0
    \end{cases} \nonumber\\
    &\qquad\times \begin{cases}
        \mathrm{e}^{2\pi i (t_\R-m_\R)\st{\frac{n_{43}}{c}}+2\pi i (\tilde{t}_\R-\tilde{m}_\R) \st{\frac{d n_{43}}{c}}}\ , & n_{43} \ne 0 \\
        \frac{(-1)^{m_\R+\tilde{m}_\R}(\mathrm{e}^{\pi i (t_\R-\tilde{t}_\R)}+\mathrm{e}^{\pi i (\tilde{t}_\R-t_\R)}-\mathrm{e}^{\pi i (t_\R+\tilde{t}_\R)}-\mathrm{e}^{-\pi i (2s+t_\R+\tilde{t}_\R)})}{ 1-\mathrm{e}^{-2\pi i s}}\ , & n_{43}=0\ .
    \end{cases} \label{eq:A na ma tilde ma final answer}
\end{align}
We used the short hand $P^{(2)}=P_{m_\D,m_\U,\tilde{m}_\D,\tilde{m}_\U}(t_\L,t_\R,\tilde{t}_\L,\tilde{t}_\R)$ given in \eqref{eq:P2 polynomial}.
This of course still looks complicated, but we should notice that the number of integrals at this point is reduced. We will be able to further simplify the expressions in the following.
\subsection{Partial sum over the masses}
We can perform the sum over $m_\L$, $\tilde{m}_\L$, $m_\R$ and $\tilde{m}_\R$ in \eqref{eq:sum over all m and mtilde}. First, we should notice that the phase prefactors in \eqref{eq:sum over all m and mtilde} in the second and third line involving $m_\L$, $m_\R$, $\tilde{m}_\L$ and $\tilde{m}_\R$ cancel the corresponding factors in \eqref{eq:A na ma tilde ma final answer} and only leave $(-1)^{m_\L+\tilde{m}_\L+m_\R+\tilde{m}_\R}$, which then factors out of the case distinctions. Thus, the sum over $m_\L$ and $m_\R$ only involves the terms
\begin{align}
&Q_{m_\D,m_\U} \frac{\Gamma(-t_\L) \Gamma(s+t_\L-m_\D-m_\U)\Gamma(-t_\R) \Gamma(s+t_\R-m_\D-m_\U)}{\Gamma(s)^2} \nonumber\\
&\qquad\equiv \sum_{m_\L,m_\R\ge 0} Q_{m_\L,m_\D,m_\R,m_\U}(-1)^{m_\L+m_\R}\nonumber\\
&\qquad\qquad\times \frac{\Gamma(m_\L-t_\L)\Gamma(s+t_\L-m_\L)\Gamma(m_\R-t_\R)\Gamma(s+t_\R-m_\R)}{\Gamma(s)^2}\ ,
\end{align}
and similarly for the tilded quantities. This defines the polynomials $Q_{m_\D,m_\U}(s,t,t_\L,t_\R)$ which are again the same polynomials as the one appearing in the Baikov representation of the imaginary part and the Rademacher expansion of the open string \cite{Eberhardt:2023xck}.

Thus, after this partial sum, the full amplitude reads
\be 
A=\sum_{c=1}^\infty \sum_{\begin{subarray}{c} a=0 \\ (a,c)=1 \end{subarray}}^{c-1} \sum_{n_\L,n_\D,n_\R=0}^{c-1} \mathrm{e}^{2\pi i \sum_{1 \le j<i\le 4} s_{ij} \big[\frac{n_{ij}}{2c}+\sum_{m=1}^{n_{ij}} \st{\frac{md}{c}}\big]} A_{a/c}^{n_\L,n_\D,n_\R}\ , \label{eq:A as sum over windings}
\ee
with
\begin{align}
    A_{a/c}^{n_\L,n_\D,n_\R}&=\frac{2^{\frac{5}{2}}}{45s t u c^{\frac{7}{2}}}\sum_{\begin{subarray}{c}
        \sqrt{m_\D}+\sqrt{m_\U} \le \sqrt{s} \\
\sqrt{\tilde{m}_\D}+\sqrt{\tilde{m}_\U} \le \sqrt{s}
    \end{subarray}}\mathrm{e}^{\frac{2\pi i}{c} (n_{32} m_\D+n_{14} m_\U)-\frac{2\pi i a m_\U}{c}+\frac{2\pi i d}{c} (n_{32} \tilde{m}_\D+n_{14} \tilde{m}_\U)-\frac{2\pi i d \tilde{m}_\U}{c}} \nonumber\\ 
    &\qquad\times\int_{P>0} \d t_\L \, \d t_\R\, \int_{\tilde{P}>0} \d \tilde{t}_\L\, \d \tilde{t}_\R \, Q \tilde{Q}\, \mathrm{e}^{-\frac{4\pi i}{c} P^{(2)}} (P \tilde{P})^{\frac{5}{4}}\nonumber\\
    &\qquad\times\Big[\sqrt{\tfrac{\tilde{P}}{P}}\Big(J_{\frac{3}{2}}\big(\tfrac{4\pi \sqrt{P \tilde{P}}}{c}\big)-J_{\frac{7}{2}}\big(\tfrac{4\pi \sqrt{P \tilde{P}}}{c}\big)\Big)+12ic \, s(a,c) J_{\frac{5}{2}}\big(\tfrac{4\pi \sqrt{P \tilde{P}}}{c}\big)\Big] \nonumber\\
    &\qquad\times \frac{\Gamma(-t_\L) \Gamma(s+t_\L-m_\D-m_\U)\Gamma(-t_\R) \Gamma(s+t_\R-m_\D-m_\U)}{\Gamma(s)^2} \nonumber \\
    &\qquad\times \frac{\Gamma(-\tilde{t}_\L) \Gamma(s+\tilde{t}_\L-\tilde{m}_\D-\tilde{m}_\U)\Gamma(-\tilde{t}_\R) \Gamma(s+\tilde{t}_\R-\tilde{m}_\D-\tilde{m}_\U)}{\Gamma(s)^2} \nonumber \\
    &\qquad\times \begin{cases}
        \mathrm{e}^{2\pi i t_\L\st{\frac{n_{21}}{c}}+2\pi i \tilde{t}_\L\st{\frac{d n_{21}}{c}}}\ , & n_{21} \ne 0 \\
        \frac{\mathrm{e}^{\pi i (t_\L-\tilde{t}_\L)}+\mathrm{e}^{\pi i (\tilde{t}_\L-t_\L)}-\mathrm{e}^{\pi i (t_\L+\tilde{t}_\L)}-\mathrm{e}^{-\pi i (2s+t_\L+\tilde{t}_\L)}}{ 1-\mathrm{e}^{-2\pi i s}}\ , & n_{21}=0
    \end{cases} \nonumber\\
    &\qquad\times \begin{cases}
        \mathrm{e}^{2\pi i t_\R\st{\frac{n_{43}}{c}}+2\pi i \tilde{t}_\R \st{\frac{d n_{43}}{c}}}\ , & n_{43} \ne 0 \\
        \frac{\mathrm{e}^{\pi i (t_\R-\tilde{t}_\R)}+\mathrm{e}^{\pi i (\tilde{t}_\R-t_\R)}-\mathrm{e}^{\pi i (t_\R+\tilde{t}_\R)}-\mathrm{e}^{-\pi i (2s+t_\R+\tilde{t}_\R)}}{ 1-\mathrm{e}^{-2\pi i s}}\ , & n_{43}=0\ ,
    \end{cases}
\end{align}
where we wrote $Q=Q_{m_\D,m_\U}(s,t,t_\L,t_\R)$ and $\tilde{Q}=Q_{\tilde{m}_\D,\tilde{m}_\U}(s,t,\tilde{t}_\L,\tilde{t}_\R)$. Remarkably, this \emph{almost} factorizes in tilded and untilded quantities, except for the term $\mathrm{e}^{-\frac{4\pi i}{c} P^{(2)}}$ and the line involving the Bessel functions, we will discuss this further in Section~\ref{subsec:factorization}.

To follow a similar naming convention for the $n_i$'s as for the rest of the parameters, we denote
\be 
n_\L=n_{21}\ , \qquad n_\D=n_{32}\ , \qquad n_\R=n_{43}\ , \qquad n_\U=-a-1-n_{41}\ .
\ee
These parameters are to be read mod $c$. Thus we have the constraint
\be 
n_\L+n_\D+n_\R+n_\U=-a-1\ .
\ee
It is also convenient to set
\begin{subequations}
\begin{align} 
t_\L&=\frac{1}{2}(m_\D+m_\U-s)+\frac{\sqrt{\Delta}}{2 \sqrt{s}}\big(x\sqrt{-u} +y\sqrt{-t}\big)\ , \\
t_\R&=\frac{1}{2}(m_\D+m_\U-s)+\frac{\sqrt{\Delta}}{2 \sqrt{s}}\big(x\sqrt{-u}-y\sqrt{-t}\big)\ ,
\end{align} \label{eq:tL tR change of variables}%
\end{subequations}
with 
\be 
\Delta=\big(s-(\sqrt{m_\D}+\sqrt{m_\U})^2\big)\big(s-(\sqrt{m_\D}-\sqrt{m_\U})^2\big)\ ,
\ee
and similarly for the tilded quantities. This change of variables maps the ellipse $P>0$ into the unit circle.
We also define
\be 
\Delta^{(2)}=s^2-(m_\D+m_\U+\tilde{m}_\D+\tilde{m}_\U)s+(m_\D-m_\U)(\tilde{m}_\D-\tilde{m}_\U)\ ,
\ee
again a sort of cross term between $\Delta$ and $\tilde{\Delta}$. In these variables we have
\be 
P=\frac{\Delta}{4s}(1-x^2-y^2)\ ,
\ee
and similarly for $\tilde{P}$. We also have
\be 
\tilde{P}^{(2)}=\frac{2s(m_\U+\tilde{m}_\U)+\Delta^{(2)}-\sqrt{\Delta\tilde{\Delta}}(x \tilde{x}+y \tilde{y})}{4s}\ .
\ee
We can then write
\begin{align}
    &A_{a/c}^{n_\L,n_\D,n_\R}\nonumber\\
    &\quad=\sum_{\begin{subarray}{c}
        \sqrt{m_\D}+\sqrt{m_\U} \le \sqrt{s} \\
\sqrt{\tilde{m}_\D}+\sqrt{\tilde{m}_\U} \le \sqrt{s}
    \end{subarray}}\frac{(\Delta \tilde{\Delta})^{\frac{9}{4}}}{720\sqrt{2} c^{\frac{7}{2}}s^{\frac{11}{2}}}\,\mathrm{e}^{\frac{2\pi i}{c} (n_\D (m_\D+d \tilde{m}_\D)+n_\U (m_\U+d \tilde{m}_\U))-\frac{\pi i}{c s} \Delta^{(2)}} \nonumber\\ 
    &\qquad\times\int_{\DD} \d x \, \d y\, \int_{\DD} \d \tilde{x}\, \d \tilde{y} \, Q \tilde{Q}\, \mathrm{e}^{\frac{\pi i \sqrt{\Delta \tilde{\Delta}}}{s c}(x \tilde{x}+y \tilde{y})} ((1-x^2-y^2)(1-\tilde{x}^2-\tilde{y}^2))^{\frac{5}{4}}\nonumber\\
    &\qquad\times\Big[\sqrt{\tfrac{\tilde{P}}{P}}\Big(J_{\frac{3}{2}}\big(\tfrac{4\pi \sqrt{P \tilde{P}}}{c}\big)-J_{\frac{7}{2}}\big(\tfrac{4\pi \sqrt{P \tilde{P}}}{c}\big)\Big)+12ic \, s(a,c) J_{\frac{5}{2}}\big(\tfrac{4\pi \sqrt{P \tilde{P}}}{c}\big)\Big]\bigg|_{\begin{subarray}{c} P\to \frac{\Delta}{4s}(1-x^2-y^2) \\ \tilde{P}\to \frac{\tilde{\Delta}}{4s}(1-\tilde{x}^2-\tilde{y}^2) \end{subarray}} \nonumber\\
    &\qquad\times \Bigg( \frac{\Gamma(-t_\L) \Gamma(s+t_\L-m_\D-m_\U)\Gamma(-\tilde{t}_\L) \Gamma(s+\tilde{t}_\L-\tilde{m}_\D-\tilde{m}_\U)}{\Gamma(s)^2} \nonumber \\
    &\qquad\qquad\quad\times \begin{cases}
        \mathrm{e}^{2\pi i t_\L\st{\frac{n_\L}{c}}+2\pi i \tilde{t}_\L\st{\frac{d n_\L}{c}}}\ , & n_\L \ne 0 \\
        \frac{\mathrm{e}^{\pi i (t_\L-\tilde{t}_\L)}+\mathrm{e}^{\pi i (\tilde{t}_\L-t_\L)}-\mathrm{e}^{\pi i (t_\L+\tilde{t}_\L)}-\mathrm{e}^{-\pi i (2s+t_\L+\tilde{t}_\L)}}{ 1-\mathrm{e}^{-2\pi i s}}\ , & n_\L=0
    \end{cases} \Bigg) \nonumber\\
    &\qquad \times (\L \leftrightarrow \R)\ . \label{eq:final Rademacher formula}
\end{align}
The symmetry between tilded and untilded parts is now perfect. The full real part of the amplitude is obtained by plugging \eqref{eq:final Rademacher formula} into \eqref{eq:A as sum over windings}. As commented below \eqref{eq:general Rademacher formula}, the imaginary part of the amplitude may be reinstated by shifting $s(a,c) \to s(a,c)+\frac{1}{4}$.

\subsection{Checks for the phase prefactor}
Let us now revisit the phase prefactor in \eqref{eq:A as sum over windings}
\be 
\mathrm{e}^{2\pi i \sum_{1 \le j<i\le 4} s_{ij} \big[\frac{n_{ij}}{2c}+\sum_{m=1}^{n_{ij}} \st{\frac{md}{c}}\big]}\ .
\ee
Consider first the function
\be 
f_{d,c}(n)=\frac{n}{2c}+\sum_{m=1}^n \bigst{\frac{md}{c}}\ .
\ee
Notice that our definition \eqref{eq:negative sawtooth sum} for the sum in the case of negative $n$ ensures that this is a periodic function. Indeed for $n \in \{1,\dots,c-1\}$, we have
\be 
f_{d,c}(n-c)=f_{d,c}(n)
\ee
as a consequence of the property
\be 
\sum_{m=1}^{c-1} \bigst{\frac{md}{c}}=0\ ,
\ee
which follows from oddness and periodicity of $\st{x}=\st{x+1}=- \st{-x}$. Thus, we can naturally extend $f_{d,c}(n)$ to all integers $n$ by continuing it periodically mod $c$.

We observe that $f_{d,c}(n)$ enjoys the following reciprocity property:
\be 
f_{d,c}(n)=f_{a,c}(dn)\ ,
\ee
where $ad \equiv 1 \bmod c$. We checked this property directly for $c\le 1000$, but do not know of a direct proof. However, it follows in principle from our derivation, since we can map $\tau=\frac{a}{c} \to \tau=\frac{d}{c}$ by exchanging left- and right-movers on the worldsheet, see \cite[Section~2.5.3]{Baccianti:2025gll}.

The phase in the exponent can be written as
\begin{multline} 
\phi_{d,c}(n_\L,n_\D,n_\R,n_\U)=s (f_{d,c}(n_\L)+f_{d,c}(n_\R))+t (f_{d,c}(n_\L+n_\D+n_\R)+f_{d,c}(n_\D))\\
+u(f_{d,c}(n_\L+n_\D)+f_{d,c}(n_\D+n_\R))\ .
\end{multline}
The properties above imply that the phase satisfies
\be 
\phi_{d,c}(n_\L,n_\D,n_\R,n_\U)=\phi_{a,c}(dn_\L,dn_\D,dn_\R,dn_\U)\ .
\ee
Notice that this also respects the constraint $n_\L+n_\D+n_\R+n_\U\equiv-a-1$, since
\be 
dn_\L+dn_\D+d n_\R+d n_\U=-ad-d\equiv-d-1\ .
\ee
Thus the symmetry between tilded and untilded quantities is also respected by the phase prefactor. We also note that everything only depends on the residue classes of $n_a$. It should also be possible to give $f_{d,c}(n)$ a purely cohomological interpretation, similar to the Dedekind sums \cite{Atiyah_logarithm}.

\section{Special cases and applications} \label{sec:special cases}
The final formula \eqref{eq:final Rademacher formula} is very general, but unfortunately also rather complicated. We will now work out a number of situations of physical interest in which the formula simplifies.

\subsection{Forward limit}
In the forward limit where $t=0$, the final formula \eqref{eq:final Rademacher formula} simplifies somewhat. Notice that $t_\L=t_\R$ and the change of variables \eqref{eq:tL tR change of variables} is in fact independent of $y$ and similarly for $\tilde{y}$. Thus we can perform the integrals over $y$ and $\tilde y$ explicitly. This can be done with the help of the identities in Appendix~\ref{app:Bessel function identities}. For the last Bessel function appearing in \eqref{eq:final Rademacher formula} we have
\begin{align}
    &\int_{-\sqrt{1-x^2}}^{\sqrt{1-x^2}} \d y\int_{-\sqrt{1-\tilde{x}^2}}^{\sqrt{1-\tilde{x}^2}} \d \tilde{y}\ \mathrm{e}^{\frac{\pi i \sqrt{\Delta \tilde{\Delta}}}{s c}y \tilde{y}} \big((1-x^2-y^2)(1-\tilde{x}^2-\tilde{y}^2)\big)^{\frac{5}{4}} \nonumber\\
    &\hspace{7cm}\times J_{\frac{5}{2}}\Big(\tfrac{\pi \sqrt{\Delta \tilde{\Delta} (1-x^2-y^2)(1-\tilde{x}^2-\tilde{y}^2)}}{sc}\Big) \nonumber\\
    &\qquad= (1-x^2)^{\frac{7}{4}}(1-\tilde{x}^2)^{\frac{7}{4}}\int_{-1}^1 \d y \int_{-1}^1 \d \tilde{y} \ \mathrm{e}^{i a y \tilde{y}} (1-y^2)^{\frac{5}{4}}(1-\tilde{y}^2)^{\frac{5}{4}} \nonumber\\
    &\hspace{7cm}\times J_{\frac{5}{2}}\big(a (1-y^2)^{\frac{1}{2}}(1-\tilde{y}^2)^{\frac{1}{2}} \big) \\
    &\qquad= \sqrt{8\pi a}(1-x^2)^{\frac{7}{4}}(1-\tilde{x}^2)^{\frac{7}{4}}\int_{0}^1 \d y \int_{0}^1 \d \tilde{y} \  (y \tilde{y})^{\frac{1}{2}} (1-y^2)^{\frac{5}{4}}(1-\tilde{y}^2)^{\frac{5}{4}} \nonumber\\
    &\hspace{7cm}\times J_{-\frac{1}{2}}(a y \tilde{y})J_{\frac{5}{2}}\big(a (1-y^2)^{\frac{1}{2}}(1-\tilde{y}^2)^{\frac{1}{2}} \big) \\
    &=\frac{5 \pi \sqrt{s c} (1-x^2)^{\frac{3}{2}}(1-\tilde{x}^2)^{\frac{3}{2}}}{8 \sqrt{2} (\Delta \tilde{\Delta})^{\frac{1}{4}}} J_3\Big(\tfrac{\pi \sqrt{(1-x^2)(1-\tilde{x}^2) \Delta \tilde{\Delta}}}{s c}\Big)\ ,
\end{align}
where $a=\frac{\pi}{s c} \sqrt{(1-x^2)(1-\tilde{x}^2) \Delta \tilde{\Delta}}$ and we used \eqref{eq:first Bessel function identity} in the last step. We can similarly evaluate the other integrals, which are
\begin{multline}
    \int \d y \int \d \tilde{y}\ \mathrm{e}^{\frac{\pi i \sqrt{\Delta \tilde{\Delta}}}{s c} y \tilde{y}} \sqrt{\frac{\tilde{\Delta}}{\Delta}} (1-x^2-y^2)^{\frac{3}{4}}(1-\tilde{x}^2-\tilde{y}^2)^{\frac{7}{4}} J_{\nu}\Big(\tfrac{\pi \sqrt{\Delta \tilde{\Delta} (1-x^2-y^2)(1-\tilde{x}^2-\tilde{y}^2)}}{sc}\Big)\\
    =\frac{5\pi \sqrt{c s} (1-x^2)(1-\tilde{x}^2)^2 \tilde{\Delta}^{\frac{1}{4}}}{8 \sqrt{2} \Delta^{\frac{3}{4}}} J_{\nu+\frac{1}{2}} \Big(\tfrac{\pi \sqrt{(1-x^2)(1-\tilde{x}^2) \Delta \tilde{\Delta}}}{s c}\Big)
\end{multline}
for both $\nu=\frac{3}{2}$ and $\nu=\frac{7}{2}$. 

Thus the forward version of \eqref{eq:final Rademacher formula} essentially shifts the indices of the Bessel functions up by $\frac{1}{2}$ and changes the prefactor,
\begin{align}
    &A_{a/c}^{n_\L,n_\D,n_\R,n_\U,m_\D,m_\U,\tilde{m}_\D,\tilde{m}_\U}\nonumber\\
    &\quad=\frac{\pi(\Delta \tilde{\Delta})^2}{2304 c^3 s^5}\,\mathrm{e}^{\frac{2\pi i}{c} (n_\D (m_\D+d \tilde{m}_\D)+n_\U (m_\U+d \tilde{m}_\U))-\frac{\pi i}{c s} \Delta^{(2)}} \nonumber\\ 
    &\qquad\times\int_{-1}^1 \d x  \int_{-1}^1 \d \tilde{x}\, Q \tilde{Q}\, \mathrm{e}^{\frac{\pi i \sqrt{\Delta \tilde{\Delta}}}{s c}x \tilde{x}} ((1-x^2)(1-\tilde{x}^2))^{\frac{3}{2}}\nonumber\\
    &\qquad\times\Big[\sqrt{\tfrac{\tilde{\Delta}(1-\tilde{x}^2)}{\Delta(1-x^2)}}\Big(J_2\big(\tfrac{\pi \sqrt{\Delta \tilde{\Delta}(1-x^2)(1-\tilde{x}^2)}}{sc}\big)-J_{4}\big(\tfrac{\pi \sqrt{\Delta \tilde{\Delta}(1-x^2)(1-\tilde{x}^2)}}{sc}\big)\Big)\nonumber\\
    &\qquad\qquad+12ic \, s(a,c) J_{3}\big(\tfrac{\pi \sqrt{\Delta \tilde{\Delta}(1-x^2)(1-\tilde{x}^2)}}{sc}\big)\Big]\nonumber\\
    &\qquad\times \Bigg(\frac{\Gamma(-t_\L) \Gamma(s+t_\L-m_\D-m_\U)\Gamma(-\tilde{t}_\L) \Gamma(s+\tilde{t}_\L-\tilde{m}_\D-\tilde{m}_\U)}{\Gamma(s)^2} \nonumber \\
    &\qquad\qquad\quad\times \begin{cases}
        \mathrm{e}^{2\pi i t_\L\st{\frac{n_\L}{c}}+2\pi i \tilde{t}_\L\st{\frac{d n_\L}{c}}}\ , & n_\L \ne 0 \\
        \frac{\mathrm{e}^{\pi i (t_\L-\tilde{t}_\L)}+\mathrm{e}^{\pi i (\tilde{t}_\L-t_\L)}-\mathrm{e}^{\pi i (t_\L+\tilde{t}_\L)}-\mathrm{e}^{-\pi i (2s+t_\L+\tilde{t}_\L)}}{ 1-\mathrm{e}^{-2\pi i s}}\ , & n_\L=0
    \end{cases} \Bigg)\nonumber\\
    &\qquad \times (\L \leftrightarrow \R)\ , \label{eq:final Rademacher formula forward limit}
\end{align}
with $t_\L=t_\R$ and $\tilde{t}_\L=\tilde{t}_\R$ given by \eqref{eq:tL tR change of variables}.

This formula can also be derived in a more straightforward manner without first deriving the general formula. Integrating in the Gaussian as we did in Section~\ref{subsec:evaluating integrals} is not necessary and the integral over the moduli can be directly performed. Since the $\sqrt{\frac{1}{c^2}-\tau \tilde{\tau}}$ term from \eqref{eq:integrating in 1} is absent, the index of the Bessel functions is shifted by $\frac{1}{2}$.

\subsection{Imaginary part}
The imaginary part of the amplitude is simpler to evaluate because it is in principle determined by unitarity from the tree-level amplitude with two massless and two massive external states. Such a unitarity cut can be implemented in string theory and leads to the Baikov representation of the imaginary part of the amplitude \cite{Eberhardt:2022zay, Banerjee:2024ibt}
\begin{multline}
    \Im A_{\text{B}}(s,0) =\frac{16\pi s^4}{15\sqrt{stu}} \sum_{\sqrt{m_\D}+\sqrt{m_\U} \le \sqrt{s}} \int \d t_\L \, \d t_\R\ P_{m_\D,m_\U}^{\frac{5}{2}} Q_{m_\D,m_\U}^2 \\
\times \frac{\Gamma(-s)\Gamma(-t_\L)\Gamma(-u_\L)}{\Gamma(1+s)\Gamma(1+t_\L)\Gamma(1+u_\L)} \times \frac{\Gamma(-s)\Gamma(-t_\R)\Gamma(-u_\R)}{\Gamma(1+s)\Gamma(1+t_\R)\Gamma(1+u_\R)}\ ,\label{eq:imaginary part Baikov representation}
\end{multline}
without the need of the full Rademacher expansion. At this point, we have two representations of the imaginary part of the amplitude. As mentioned above, we can obtain both the real and imaginary part of the Rademacher formula \eqref{eq:final Rademacher formula} by shifting $s(a,c) \to s(a,c)+\frac{1}{4}$. Thus the imaginary part is given by replacing $s(a,c) \to \frac{1}{4}$ and omitting the $J_{\frac{3}{2}}$ and $J_{\frac{7}{2}}$ terms in \eqref{eq:final Rademacher formula}. Equality of these two expressions is highly non-trivial. As a consistency check of our analysis, we verified this equality numerically and found very good agreement, see Figure~\ref{fig:Baikov-comparison}. This tests essentially all parts of the formula.

\subsection{\label{sec:mass shifts}Mass shifts and decay widths}
The averaged mass shifts and decay widths are read off from the coefficients of double poles at $s = n \in \ZZ_{\ge 1}$, see \cite{Sundborg:1988ai,Amano:1988ht,Sundborg:1989jv,Okada:1989sd,Marcus:1988vs,Mitchell:1988qe,Chialva:2003hg,Sen:2016gqt,Eberhardt:2022zay,Eberhardt:2023xck} for previous investigations in different string theories. Let us use the notation $A = \text{DRes}_{s=n}A / (s- n)^2 + \ldots$. At the level of the formula \eqref{eq:final Rademacher formula}, these double poles come from the denominator $1-\mathrm{e}^{-2\pi i s}$ in the $n_\L=n_\R=0$ case. Therefore, we can restrict to the $n_\L=n_\R=0$ contribution for the mass shifts. This also accelerates convergence which explains why the coincidence of the Rademacher answer with the Baikov answer is perfect in Figure~\ref{fig:Baikov-comparison}.

The combination of Gamma functions in the integrand \eqref{eq:final Rademacher formula} becomes polynomial in this case. Changing to polar coordinates, one can perform the integration over the angular variables relatively straightforwardly. The final integral over the radial coordinates can be performed with the help of the identities in Appendix~\ref{app:Bessel function identities}. For example, for the mass shift at $s=1$, the relevant integrals take the form
\begin{align}
    \DRes_{s=1} A_{a/c}^{0,n_\D,0,n_\U,0,0,0,0}&=- \frac{\mathrm{e}^{-\frac{\pi i}{c}} \pi^4}{45 \sqrt{2} c^{\frac{7}{2}}} \int_0^1 \d r \int_0^1 \d \tilde{r}\ r \tilde{r} \, J_0\big(\tfrac{\pi r \tilde{r}}{c}\big)\, \big((1-r^2)(1-\tilde{r}^2)\big)^{\frac{5}{4}}  \nonumber\\
    &\qquad\times\Big[\sqrt{\tfrac{1-\tilde{r}^2}{1-r^2}}\Big(J_{\frac{3}{2}}\big(\tfrac{\pi \sqrt{(1-r^2)(1-\tilde{r}^2)}}{c}\big)-J_{\frac{7}{2}}\big(\tfrac{\pi \sqrt{(1-r^2)(1-\tilde{r}^2)}}{c}\big)\Big)\nonumber\\
    &\qquad\qquad+12ic \, s(a,c) J_{\frac{5}{2}}\big(\tfrac{\pi \sqrt{(1-r^2)(1-\tilde{r}^2)}}{c}\big)\Big]\\
    &=-\frac{\mathrm{e}^{-\frac{\pi i}{c}} \sqrt{2}\pi^3}{630 c^{\frac{5}{2}}} \Big[ J_{\frac{5}{2}}\big(\tfrac{\pi}{c}\big)-J_{\frac{9}{2}}\big(\tfrac{\pi}{c}\big)+12 i c \, s(a,c) J_{\frac{7}{2}}\big(\tfrac{\pi}{c}\big)\Big]\ .
\end{align}
When we combine this with the phase prefactor present in \eqref{eq:A as sum over windings}, we get the relatively simple expression
\be 
\DRes_{s=1} A_{a/c}=-\sum_{n=0}^{c-1} \frac{\mathrm{e}^{-\frac{\pi i}{c}(2d n^2+2(d+1)n+1)} \sqrt{2}\pi^3}{630 c^{\frac{5}{2}}} \Big[ J_{\frac{5}{2}}\big(\tfrac{\pi}{c}\big)-J_{\frac{9}{2}}\big(\tfrac{\pi}{c}\big)+12 i c \, s(a,c) J_{\frac{7}{2}}\big(\tfrac{\pi}{c}\big)\Big]\ ,
\ee
which then has to be summed over $a$ and $c$ to get the full mass shift. This then recovers the expression for the mass shift discussed in \cite[Section~3.3]{Baccianti:2025gll}, where it was directly derived from a Rademacher expansion of the modular integral after integration over the positions of punctures.\footnote{To make the expressions literally agree we have to rename $d \to a$ and $n \to -m-1$.}

We can similarly produce analytic formulas for higher mass shifts -- these will be polynomials of order $2s-2$, which could then be decomposed into Gegenbauer polynomials to extract the mass shift of a given higher spin component. The formulas quickly get very unwieldy. We have for example
\begin{align}
    \DRes_{s=2} A_{a/c}&=-\frac{\pi ^3 \big(\mathrm{e}^{-\frac{i \pi  (4 a+3)}{2 c}} G(-2 d,d+3,c)+\mathrm{e}^{-\frac{i \pi }{2 c}} G(-2 d,d+1,c)\big)}{10478160
   c^{\frac{5}{2}}} \nonumber\\
   &\qquad\times \Big(1584 (t+1)^2 J_{\frac{5}{2}}(\tfrac{\pi }{2 c})+19712 i c s(a,c) (t+1)^2 J_{\frac{7}{2}}(\tfrac{\pi }{2
   c})\nonumber\\
   &\qquad\quad-3 (585 t^2+1170 t+592) J_{\frac{9}{2}}(\tfrac{\pi }{2 c})\nonumber\\
   &\qquad\quad-28 i c s(a,c) (7 t^2+14 t+16) J_{\frac{11}{2}}(\tfrac{\pi }{2 c})-21 t (t+2)
   J_{\frac{13}{2}}(\tfrac{\pi }{2 c})\Big) \nonumber\\
   &\quad-\frac{\pi^3\big(\mathrm{e}^{-\frac{i \pi  (2 a+3)}{c}} G(-2 d,2 d+3,c)+\mathrm{e}^{-\frac{i \pi }{c}} G(-2 d,2 d+1,c)\big)}{1309770 \sqrt{2}
   c^{\frac{5}{2}}} \nonumber\\
   &\qquad\times\Big(198 (t+1)^2 J_{\frac{5}{2}}(\tfrac{\pi }{c})+1232 i c s(a,c) (t+1)^2 J_{\frac{7}{2}}(\tfrac{\pi
   }{c})\nonumber\\
   &\qquad\quad-3 (53 t^2+106 t+46) J_{\frac{9}{2}}(\tfrac{\pi }{c})+14 i c s(a,c) (7 t^2+14 t+16) J_{\frac{11}{2}}(\tfrac{\pi }{c})\nonumber\\
   &\qquad\quad+21 t (t+2)
   J_{\frac{13}{2}}(\tfrac{\pi }{c})\Big) \nonumber\\
   &\quad-\frac{\pi ^3 \mathrm{e}^{-\frac{i \pi }{c}} G(-2 d,d+2,c)}{2619540 \sqrt{2} c^{\frac{5}{2}}} \Big(264 (t+1)^2 J_{\frac{5}{2}}(\tfrac{\pi }{c})+4928 i c s(a,c) (t+1)^2
   J_{\frac{7}{2}}(\tfrac{\pi }{c})\nonumber\\
   &\qquad\quad-3 (45 t^2+90 t+38) J_{\frac{9}{2}}(\tfrac{\pi }{c})+56 i c s(a,c) (7 t^2+14 t+16) J_{\frac{11}{2}}(\tfrac{\pi }{c})\nonumber\\
   &\qquad\quad-21 (t^2+2 t+2)
   J_{\frac{13}{2}}(\tfrac{\pi }{c})\Big)\nonumber\\
   &\quad-\frac{4 \pi^3 \mathrm{e}^{-\frac{2 i \pi
   }{c}} G(-2 d,2 d+2,c)}{654885 c^{\frac{5}{2}}} \Big(33
   (t+1)^2 J_{\frac{5}{2}}(\tfrac{2 \pi }{c})\nonumber\\
   &\qquad\quad+308 i c s(a,c) (t+1)^2 J_{\frac{7}{2}}(\tfrac{2 \pi
   }{c})-3 (10 t^2+20 t+17) J_{\frac{9}{2}}(\tfrac{2 \pi }{c})\nonumber\\
   &\qquad\quad-28 i c s(a,c) (7 t^2+14 t+16) J_{\frac{11}{2}}(\tfrac{2 \pi }{c})+21 (t^2+2 t+2) J_{\frac{13}{2}}(\tfrac{2 \pi }{c})\Big)\ .
\end{align}
We denoted by
\begin{align}
    G(a,b,c)=\sum_{n=0}^{c-1} \mathrm{e}^{\frac{2\pi i (a n^2+bn)}{c}}
\end{align}
the general Gauss sum, which can be efficiently computed using Gauss' evaluation in terms of Legendre symbols. We systematically worked out analogous formulas for higher mass shifts \cite{BEM}. Imaginary parts can again be determined exactly from the unitarity formula \eqref{eq:imaginary part Baikov representation} and we have for example simply
\begin{subequations}
\begin{align}
    \Im \DRes_{s=1} A&=\frac{\pi ^2}{210}\ , \\
    \Im \DRes_{s=2} A&=\frac{\pi ^2 (207 t^2+414 t+272)}{166320}\ , \\
    \Im \DRes_{s=3} A&=\frac{\pi ^2 (537893 t^4+3227358 t^3+7403607 t^2+7687710 t+3261735)}{3502699200}\ .
\end{align}
\end{subequations}
Starting from $s=5$, expressions also contain square roots since $\Delta_{m_\D=1,m_\U=1}(s=5)=5$ is no longer a square. 
Thus the imaginary parts of mass shifts (divided by $\pi^2$ in our conventions) take in general values in
\be 
\pi^{-2}\Im \DRes_{s} A  \in \QQ\bigg[t,\bigcup_{\sqrt{m_\D}+\sqrt{m_\U} \le \sqrt{s}} \sqrt{\Delta_{m_\D,m_\U}(s)}\bigg] \ .
\ee
We do not know whether the real parts are periods or any other specific transcendentality properties.

\begin{figure}
    \centering
    \includegraphics[width=\textwidth]{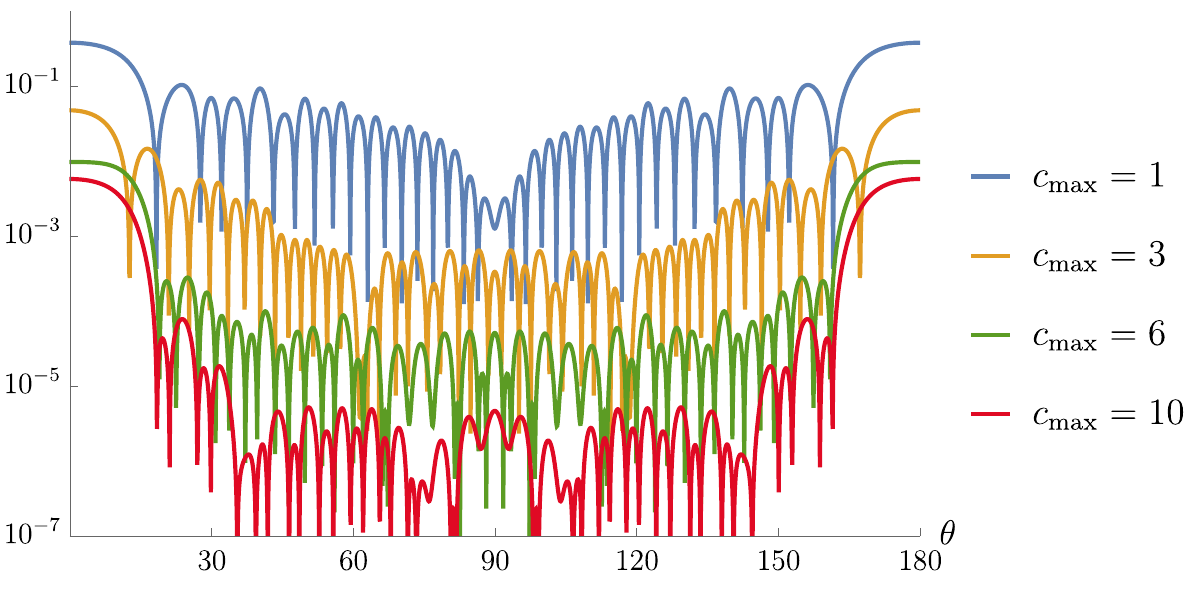}
    \caption{The relative difference $|\frac{1}{A_\text{B}}\sum_{c=1}^{c_\text{max}}\Im A_{c}-1|$ of the Rademacher formula with cutoff $c_\text{max}$ and the Baikov representation $A_\text{B}$ of the imaginary part at $s=35$ as a function of the angle.}
    \label{fig:massshift convergence 35}
\end{figure}

We checked again numerically that the imaginary parts as computed from the Rademacher expressions match the imaginary parts from the Baikov representation. This match is numerically much better than the general amplitude since the Rademacher expansion converges faster for the mass shifts than for the general amplitude since there are less sums over the windings to perform, but the large-$c$ behavior of the summands remains the same. For illustration, we plotted the relative difference of the imaginary part of the residue at $s=35$ as computed from the Baikov formula and computed from the Rademacher expansion as a function of the angle in Figure~\ref{fig:massshift convergence 35}. We see that already for $c_\text{max}=10$, the relative error is already $\sim 0.0001\%$ for a large range of angles. Convergence in the forward limit is worse since the phases that appear in the Rademacher formula tend to be less random and there are fewer cancellations.

\subsection{Large-\texorpdfstring{$c$}{c} factorization and comparison to the open string} \label{subsec:factorization}
We derived a very similar Rademacher expansion of the open string amplitudes in type-I string theory previously \cite{Eberhardt:2023xck}. It takes a very similar form. Let us for simplicity focus on $s$-channel scattering for the color-ordered planar amplitude (receiving contributions from the annulus and the M\"obius strip). It takes the form
\begin{align}
    \Re A_\text{open}=\sum_{c=1}^\infty \sum_{\begin{subarray}{c} 0 \le a \le \frac{c}{2} \\ (a,c)=1 \end{subarray}} \frac{2-\delta_{a,0}-\delta_{a,\frac{c}{2}}}{2} \hspace{-.4cm} \sum_{\begin{subarray}{c} n_\L,n_\D,n_\R,n_\U \ge 0 \\
n_\L+n_\D+n_\R+n_\U=c-1
\end{subarray}} \hspace{-.4cm}\mathrm{e}^{2\pi i \sum_{i>j} s_{ij} \sum_{m=1}^{n_{ij}} \st{\frac{md}{c}}}A^{n_\L,n_\D,n_\R}_{\text{open},a/c}\ ,
\end{align}
with $n_\L=n_{21}$, $n_\D=n_{32}$, $n_\R=n_{43}$ and
\begin{align}
    A_{\text{open},a/c}^{n_\L,n_\D,n_\R}&=-\frac{16\pi i}{15c^5 \sqrt{stu}} \sum_{\begin{subarray}{c} m_\D,m_\U \ge 0 \\
    (\sqrt{m_\D}+\sqrt{m_\U})^2 \le s \end{subarray}}\mathrm{e}^{\frac{2\pi i d}{c}(m_\D n_\D+m_\U n_\U)}\int_{P_{m_\D,m_\U} > 0} \!\!\!\!\!\!\!\!\!\!\!\!\!\!\!\! \d t_\L \, \d t_\R\ P_{m_\D,m_\U}^{\frac{5}{2}}\, Q_{m_\D,m_\U} \nonumber\\
    &\times \left( \frac{\Gamma(-t_\L)\Gamma(s+t_\L-m_\D-m_\U)}{\Gamma(s)}\begin{cases}\mathrm{e}^{2\pi i t_\L \st{\frac{d n_\L}{c}}} &\text{if}\;\; n_\L>0 \\
    \frac{\sin(\pi(s+t_\L))}{\sin(\pi s)} &\text{if}\;\; n_\L=0
    \end{cases} \right) \big(\L \leftrightarrow \R\big)\, . 
\end{align}
The ranges of the summation are slightly different which has to do with the specifics of the moduli space of annuli and the fact that vertex operators on the boundary of the annulus are ordered. In many ways, this formula behaves like the square root of the closed string formula \eqref{eq:final Rademacher formula} before we sum over $a$, $c$ and the windings $n_a$, which can be viewed as a kind of a double-copy or a KLT formula at one loop, see also \cite{Casali:2020knc,Stieberger:2022lss,Stieberger:2023nol,Mazloumi:2024wys}. However, this factorization is of course not literally true because of the presence of the Bessel functions. We can however observe that the closed string summand factorizes to leading order in the large $c$ expansion. For large $c$, the dominant contribution comes from the Bessel function $J_{\frac{5}{2}}$ in \eqref{eq:final Rademacher formula}, since $s(a,c)$ grows on average like $\log^2(c)$.\footnote{More precisely, $\frac{1}{4\pi^2} \varphi(c) \log^2(c) \le \sum_{a=0}^{c-1} |s(a,c)| \le \frac{1}{2\pi^2} \varphi(c) \log^2(c)$ with $\varphi(c)$ the Euler totient function. For special choices of $a$, $s(a,c)$ can however be much bigger or smaller than this typical value \cite{KurtGirstmair2005}. \label{footnote:size s(a,c)}} The closed string formula exactly factorizes into the open string formula, including all the phase prefactors. Assuming that $n_\L \ne 0$ and $n_\R \ne 0$ for simplicity, we have
\begin{align}
    A_{\text{closed},a/c}^{n_\L,n_\D,n_\R} \overset{c \to \infty}{\longrightarrow}-4i s(a,c) c^5\,  A_{\text{open},a/c}^{n_\L,n_\D,n_\R}A_{\text{open},d/c}^{dn_\L,dn_\D,dn_\R}\ ,
\end{align}
with slightly more complicated formulas holding in the case $n_\L=0$ or $n_\R=0$. Thus we can view $A_{\text{open},a/c}^{n_\L,n_\D,n_\R}$ as the building blocks of a KLT formula and the Rademacher procedure expresses the closed string amplitude as a double copy of these building blocks. However, this is of limited value as we do not know of a useful extension away from the asymptotic regions of $c$.

\subsection{\label{sec:low energy expansion}Low energy expansion and convergence}
The Rademacher formula makes all branch cuts due to massive thresholds manifest, but this is unfortunately not the case for the massless branch cut. Precisely for $s \to 0$, the convergence of the Rademacher formula actually breaks down. For large $c$, $A_{a/c}^{n_\L,n_\D, n_\R}$ is of order $c^{-5} s(a,c)$. $s(a,c)$ is typically of order $\log^2(c)$, see footnote \ref{footnote:size s(a,c)}. Thus in the worst case, after performing the summations over $n_a$ and $a$, we get expressions that are of order $c^{-1}\log^2(c)$ and the sum over $c$ actually diverges. For any finite $s$, this actually does not happen because the phases behave sufficiently randomly to make the sum over $n_a$ and $a$ much smaller and the sum over $c$ convergent. However, in the low energy limit, all phases go to unity which manifests itself as the massless branch cut in the resummed expression.

The low energy expansion of the amplitude is known from other methods \cite{Green:2008uj, DHoker:2019blr,Claasen:2024ssh}. In our conventions, the amplitude of those papers should be divided by $\frac{1}{2\pi}$ to match with our conventions. In particular, we manifestly have $\lim_{s,t \to 0} A(s,t)= \text{vol}(\mathcal{F})=\frac{\pi}{3}$ in our conventions. These papers derive
\begin{align}
    A(s,t)=\frac{\pi}{3} \Big(&A_\text{sugra}+A_4+A_6+A_7\\
    &+1+\frac{\zeta_3 \sigma_3}{3}+\frac{29 \zeta_5 \sigma_2 \sigma_3}{180}+\frac{\zeta_3^2 \sigma_3^2}{18}-\frac{163 \zeta_7 \sigma_2^2 \sigma_3}{30240}+ \dots\Big)\nonumber
\end{align}
with
\begin{subequations}
\begin{align}
    A_\text{sugra}&=\frac{2}{5}\big[ I(s,t)+I(u,t)+I(s,u)\big]\ , \\
    A_4&=-\frac{4 \zeta_3}{15} (s^4 \log(-s)+\text{perm})+\frac{2 \zeta_3 \sigma_2^2}{15}\Big(-\frac{\zeta'_{-3}}{\zeta_{-3}}-\frac{\zeta'_3}{\zeta_3}+\frac{79}{60}\Big)\, , \\
    A_6&=-\frac{84 s^6+2s^4 \sigma_2}{420} \zeta_5 \log(-s)+\text{perm} \nonumber\\
    &\qquad+\frac{\zeta_5 \sigma_2^3}{630} \Big(\frac{35\zeta'_{-1}}{\zeta_{-1}}- \frac{70\zeta'_{-3}}{\zeta_{-3}}+\frac{2 \zeta'_{-5}}{\zeta_{-5}}-\frac{33 \zeta'_{5}}{\zeta_5}+\frac{13487}{420}\Big) \ , \\
    A_7&=\frac{-26 s^7+s^5 \sigma_2}{210} \zeta_3^2 \log(-s)+\text{perm}\nonumber\\
    &\qquad+\frac{\zeta_3^2 \sigma_2^2 \sigma_3}{630} \Big(- \frac{28\zeta'_{-3}}{\zeta_{-3}}- \frac{15\zeta'_{-5}}{\zeta_{-5}}-\frac{58 \zeta'_3}{\zeta_3}+15 \gamma+\frac{50417}{1260}\Big)\ ,
\end{align}
\end{subequations}
and
\be 
I(s,t)=\frac{u}{4}+\frac{st}{2u}+\frac{s^2(s+3t)\log(-s)+t^2 (t+3s) \log(-t)}{2u^2}-\frac{s^2 t^2}{u^3}\Big[L\Big(-\frac{u}{t}\Big)+L\Big(-\frac{u}{s}\Big)\Big]\ ,
\ee
and 
\be 
L(x) := \text{Li}_2(x)+\log(x)\log(1-x)=\frac{\pi^2}{6}-\text{Li}_2(1-x)\ .
\ee
Furthermore, we used the abbreviations $\zeta_n=\zeta(n)$ for Riemann zeta function evaluated at integers as well as $\zeta_n'$ for its derivative, $\sigma_n=s^n+t^n+u^n$, and $\gamma$ is the Euler--Mascheroni constant. Let us also mention that the imaginary part of the $\alpha'$-expansion can be computed to arbitrary order in $\alpha'$ since it is given by the known unitarity cut formula \eqref{eq:imaginary part Baikov representation}, see \cite[Section~6.7]{Eberhardt:2022zay} for an algorithm and \cite{DHoker:2019blr,Edison:2021ebi,Green:1999pv,Green:2008uj,DHoker:2015gmr} for previous results.

\section{\label{sec:conclusions}Conclusions}

In this paper, we synthesized the tools developed in our previous works \cite{Eberhardt:2022zay,Eberhardt:2023xck,Eberhardt:2024twy,Banerjee:2024ibt,Baccianti:2025gll} to evaluate the one-loop four-graviton scattering amplitude $\A_1(s,t)$ in type-II string theory exactly in $\alpha'$. The main result is a new representation of $\A_1(s,t)$ as an infinite sum given explicitly in Equation~\eqref{eq:A as sum over windings} and \eqref{eq:final Rademacher formula}. It can be used in practice to evaluate the amplitude at any finite value of $s$ and $t$ and it was used to produce plots in Figure~\ref{fig:intro-fixed-angle} and \ref{fig:intro-Regge}. 
This representation was cross-checked against many physical expectations,  including agreement with unitarity cuts.

Our approach paves a clear path towards generalizations to other types of string theories, such as heterotic string, and other kinds of external states. Together with the Lorentzian contours developed in \cite{Eberhardt:2024twy}, it can be potentially extended to higher multiplicity. Likewise, an interesting problem would be to extend this analysis to mixed open-closed scattering amplitudes at one loop \cite{Stieberger:2021daa}. We also think that this formalism is useful to compute higher-loop corrections in a non-perturbative instanton sector following \cite{Sen:2021tpp, Agmon:2022vdj}.

The most important outcome of this work is a numerical evaluation of the amplitude $\A_1(s,t)$ in the high-energy fixed-angle limit, i.e., both $s$ and $t$ large with their ratio fixed. It is illustrated in Figure~\ref{fig:intro-fixed-angle} for $t = -\tfrac{s}{4}$. Just like in the case of open strings \cite{Eberhardt:2023xck}, this result is in tension with the anticipated Gross--Mende behavior \cite{Gross:1987ar,Gross:1987kza}. On the one hand, the exponential envelope seems to be in agreement, but on the other, the oscillations on top of it indicate presence of additional saddle points with the same real part of the action but different phases and Hessians. Indeed, we believe that such saddles exist on the analytic continuation of the moduli space $\M^{\mathbb{C}}_{1,4}$. Investigation of this intriguing possibility will be a subject of an upcoming publication \cite{BEM}.

\section*{Acknowledgments}
We thank Guillaume Bossard, Johannes Br\"odel, Jeevan Chandra, Emiel Claasen, Mehregan Doroudiani, David Gross, Kelian H\"aring, Tom Hartman, Jan Manschot, Boris Pioline and Stephan Stieberger for useful discussions. 
MMB and LE are supported by the European Research Council (ERC) under the European Union’s Horizon 2020 research and innovation programme (grant agreement No 101115511). 

\appendix

\section{Bessel function identities} \label{app:Bessel function identities}
In this Appendix we work out several integral identities of Bessel functions which we have not found in the literature. They are needed to evaluate the remaining integrals in the Rademacher formula in special cases.
\subsection{Basic identity}
We have the basic identity
\begin{multline}
    \int_0^1 \d x \, \d y\ (xy)^{\mu+1} \big((1-x^2)(1-y^2)\big)^{\frac{\nu}{2}} J_\mu(a x y) J_\nu \big(a (1-x^2)^{\frac{1}{2}}(1-y^2)^{\frac{1}{2}}\big) \\
    =\frac{\Gamma(\mu+1)\Gamma(\nu+1)}{2a\, \Gamma(\mu+\nu+2)}\, J_{\mu+\nu+1}(a)\ .
\end{multline}
To prove this, we write out the series expansion of the Bessel function and integrate term by term. Each integral in both $x$ and $y$ is a standard beta-function integral. This exhibits the integral as an infinite double sum over of the following form
\be 
\sum_{m,n=0}^\infty \frac{(-1)^{m+n} \Gamma (m+\mu +1) \Gamma (n+\nu +1) }{4 m! n! \, \Gamma (m+n+\mu +\nu +2)^2} \Big(\frac{a}{2}\Big)^{\mu +2 m+\nu +2 n}\ .
\ee
We can reorganize the sum by summing over $k=m+n$ and $n=0,\dots,k$ instead. The sum over $n$ is of hypergeometric form and can be done with the help of the following summation formula (following from the Gauss evaluation of ${}_2F_1(a,b,c;1)$),
\begin{align}
    \sum_{n=0}^k \frac{\Gamma(k-n+\mu+1)\Gamma(n+\nu+1)}{(k-n)!n!}=\frac{\Gamma(\mu+1)\Gamma(\nu+1)\Gamma(\mu+\nu+k+2)}{\Gamma(k+1)\Gamma(\mu+\nu+2)}\ .
\end{align}
This evaluates the integral as
\begin{align}
    \frac{\Gamma(\mu+1)\Gamma(\nu+1)}{\Gamma(\mu+\nu+2)} \sum_{k=0}^\infty \frac{(-1)^k (\frac{a}{2})^{\mu +\nu +2 k}}{4k!\, \Gamma(\mu+\nu+k+2)}=\frac{\Gamma(\mu+1)\Gamma(\nu+1)}{2a\, \Gamma(\mu+\nu+2)} J_{\mu+\nu+1}(a)\ ,
\end{align}
as claimed.
\subsection{Generalizations}
We will also need generalizations of this basic formula. We now consider larger exponents of $x$ and $y$. We claim that for $p \in \ZZ_{\ge 0}$, $q \in \ZZ_{\ge 0}$,
\begin{multline}
    \int_0^1 \d x \, \d y\ x^{\mu+2p+1} y^{\mu+2q+1} \big((1-x^2)(1-y^2)\big)^{\frac{\nu}{2}} J_\mu(a x y) J_\nu \big(a (1-x^2)^{\frac{1}{2}}(1-y^2)^{\frac{1}{2}}\big) \\
    =\sum_{m=0}^{\min(p,q)} \frac{(-1)^m \Gamma(\mu+p+q+1-m) \Gamma(\nu+m+1)\Gamma(p+1)\Gamma(q+1)}{4\, \Gamma(p+1-m)\Gamma(q+1-m)\Gamma(\mu+\nu+p+q+2)\Gamma(m+1)} \\
    \times \Big(\frac{a}{2}\Big)^{-m-1} J_{\mu+\nu+m+1}(a)\ . \label{eq:first Bessel function identity}
\end{multline}
Because of the recurrence relations of Bessel functions,
\be 
J_{\nu+1}(a)+J_{\nu-1}(a)=\frac{2\nu}{a} J_\nu(a)\ ,
\ee
there are many different ways to express the right hand side as linear combinations of Bessel functions.

To prove this, we proceed in the same way as before. We simply expand the Bessel functions and integrate term by term, after which we can group terms with the same exponent of $a$ together and perform the sum. This writes the integral in the form
\begin{multline}
    \sum_{k=0}^\infty\frac{(-1)^k \Gamma(k+\mu+p+1)\Gamma(k+\mu+q+1)\Gamma(\nu+1)}{4 k! \Gamma(k+\mu+1)\Gamma(k+\mu+\nu+p+2)\Gamma(k+\mu+\nu+q+2)}\Big(\frac{a}{2}\Big)^{\mu+\nu+2k} \\
    \times {}_3 F_2 \bigg[\begin{matrix}
        -k \ \  -k-\mu \ \ \nu+1 \\ -k-p-\mu \ \  -k-q-\mu 
    \end{matrix}; 1 \bigg]\ . \label{eq:first Bessel function identity fixed a exponents 1}
\end{multline}
To continue, we need various identities of the generalized hypergeometric function ${}_3F_2$ at unit argument. Besides permutation of the first three indices and the last two indices, it also satisfies the Thomae formula and the three-term relation \cite{Bailey:1935}
\begin{subequations}
\begin{align}
{}_3F_2\bigg[ \begin{matrix} a\ \ b\ \ c \\ d\ \ e\end{matrix};1\bigg]&=\frac{\Gamma(e)\Gamma(d+e-a-b-c)}{\Gamma(e-a)\Gamma(d+e-b-c)} \, {}_3F_2\bigg[\begin{matrix} a \ \ d-b\ \ d-c \\ d\ \ d+e-b-c \end{matrix};1\bigg]\ , \\
{}_3F_2\bigg[ \begin{matrix} a\ \ b\ \ c \\ d\ \ e\end{matrix};1\bigg]&=\frac{\Gamma(1-a)\Gamma(d)\Gamma(e)\Gamma(c-b)}{\Gamma(d-b)\Gamma(e-b)\Gamma(b-a+1)\Gamma(c)} \, {}_3F_2\bigg[\begin{matrix} b \ \ b-d+1\ \ b-e+1 \\ 1+b-c\ \ 1+b-a \end{matrix};1\bigg]\nonumber\\
&\qquad+(b \leftrightarrow c)\ .
\end{align}
\end{subequations}
We will use the second relation for $b \in \ZZ_{\ge 0}$, so that the Gamma function $\Gamma(c)$ in the denominator makes the second term vanish identically. We can iterate these identities and find many relations between hypergeometric functions. The identity that we will need is, for $b \in \ZZ_{\le 0}$ or $a \in \ZZ_{\le 0}$
\begin{multline}
    {}_3F_2\bigg[ \begin{matrix} a\ \ b\ \ c \\ d\ \ e\end{matrix};1\bigg]=\frac{\Gamma(d)\Gamma(a-e+1)\Gamma(b-e+1)\Gamma(c-e+1)}{\Gamma(d-c)\Gamma(1-e)\Gamma(1+a+c-e)\Gamma(1+b+c-e)}\\
    \times {}_3F_2\bigg[ \begin{matrix} a+b+c-d-e+1 \ \  c-e+1\ \ c \\ a+c-e+1\ \ b+c-e+1\end{matrix};1\bigg]\ .
\end{multline}
We then plug in $a=-p$, $b=-q$, $c=\nu+1$, $d=-p-q-\mu$ and $e=k+2+\mu+\nu$, which turns the hypergeometric function appearing on the right hand side into the hypergeometric function appearing in \eqref{eq:first Bessel function identity fixed a exponents 1}. Using this identity, we can express the integral also as
\begin{multline}
    \sum_{k \ge 0} \frac{\Gamma(\nu+1)\Gamma(p+q+\mu+1)(-1)^k}{4k! \Gamma(\mu+\nu+k+2)\Gamma(p+q+\mu+\nu+2)} \Big(\frac{a}{2}\Big)^{\mu+\nu+2k} \\
    \times {}_3 F_2 \bigg[\begin{matrix}
        -p \ \  -q \ \ \nu+1 \\ -p-q-\mu \ \  k+\mu+\nu+2 
    \end{matrix}; 1 \bigg]\ .
\end{multline}
Finally, we insert the series definition of the hypergeometric function, 
\be 
{}_3F_2\bigg[ \begin{matrix} a\ \ b\ \ c \\ d\ \ e\end{matrix};1\bigg]= \sum_{m \ge 0} \frac{(a)_m(b)_m(c)_m}{(d)_m (e)_m\, m!}\ .
\ee
The sum over $k$ can be performed for each term and recovers the definition of the Bessel function. This shows \eqref{eq:first Bessel function identity}.

\subsection{Another identity}
We also need a second formula, where an extra $(1-x^2)^{-1}$ is inserted, which we expand in terms of a geometric series. This yields
\begin{align}
     &\int_0^1 \d x \, \d y\ x^{\mu+2p+1} y^{\mu+2q+1} (1-x^2)^{\frac{\nu}{2}-1}(1-y^2)^{\frac{\nu}{2}} J_\mu(a x y) J_\nu \big(a (1-x^2)^{\frac{1}{2}}(1-y^2)^{\frac{1}{2}}\big) \nonumber\\
    &\qquad=\sum_{\ell=0}^\infty \int_0^1 \d x \, \d y\ x^{\mu+2(p+\ell)+1} y^{\mu+2q+1} \big((1-x^2)(1-y^2)\big)^{\frac{\nu}{2}} \nonumber\\
    &\qquad\qquad\times J_\mu(a x y) J_\nu \big(a (1-x^2)^{\frac{1}{2}}(1-y^2)^{\frac{1}{2}}\big) \\ &\qquad=\sum_{\ell=0}^\infty \sum_{m=0}^{q} \frac{(-1)^m \Gamma(\mu+p+\ell+q+1-m) \Gamma(\nu+m+1)\Gamma(p+\ell+1)\Gamma(q+1)}{4\, \Gamma(p+\ell+1-m)\Gamma(q+1-m)\Gamma(\mu+\nu+p+\ell+q+2)\Gamma(m+1)} \nonumber\\
    &\qquad\qquad\times \Big(\frac{a}{2}\Big)^{-m-1} J_{\mu+\nu+m+1}(a) \\
    &\qquad=\sum_{m=0}^q \frac{(-1)^m \Gamma(p+1)\Gamma(q+1) \Gamma(p+q+\mu-m+1)\Gamma(m+\nu+1)}{4 \Gamma(m+1)\Gamma(1-m+p)\Gamma(1-m+q)\Gamma(p+q+\mu+\nu+2)} \Big(\frac{a}{2}\Big)^{-m-1} \nonumber\\
    &\qquad\qquad\times {}_3 F_2 \bigg[ \begin{matrix} 1\ \ p+1\ \ 1-m+p+q+\mu \\ 1-m+p \ \ p+q+\mu+\nu+2\end{matrix};1\bigg] J_{\mu+\nu+m+1}(a)\ . 
\end{align}
The Thomae relations for ${}_3F_2$ give the transformation
\be 
{}_3 F_2 \bigg[ \begin{matrix} 1\ \ p+1\ \ 1-m+p+q+\mu \\ 1-m+p \ \ p+q+\mu+\nu+2\end{matrix};1\bigg]=\frac{\mu+\nu+p+q+1}{\nu} {}_3 F_2 \bigg[ \begin{matrix} -m\ \ 1\ \ -q-\mu \\ 1-m+p \ \ \nu+1\end{matrix};1\bigg]\ ,
\ee
which then yields the final form of the identity
\begin{multline}
    \int_0^1 \d x \, \d y\ x^{\mu+2p+1} y^{\mu+2q+1} (1-x^2)^{\frac{\nu}{2}-1}(1-y^2)^{\frac{\nu}{2}} J_\mu(a x y) J_\nu \big(a (1-x^2)^{\frac{1}{2}}(1-y^2)^{\frac{1}{2}}\big) \\
    =\sum_{m=0}^q\frac{(-1)^m \Gamma(p+1)\Gamma(q+1) \Gamma(p+q+\mu-m+1)\Gamma(\nu)\Gamma(m+\nu+1)}{4 \Gamma(m+1)\Gamma(1-m+q)\Gamma(p+q+\mu+\nu+1)} \\
    \times \Big(\frac{a}{2}\Big)^{-m-1}{}_3 \tilde{F}_2 \bigg[ \begin{matrix} -m\ \ 1\ \ -q-\mu \\ 1-m+p \ \ \nu+1\end{matrix};1\bigg]J_{\mu+\nu+m+1}(a)\ .  \label{eq:second Bessel function identity}
\end{multline}
Notice that since $m$ is a positive integer, the series expansion of the hypergeometric function always truncates. We wrote the result in terms of the regularized hypergeometric function
\be 
{}_3 \tilde{F}_2 \bigg[ \begin{matrix} a\ \ b\ \ c \\ d \ \ e\end{matrix};1\bigg]=\frac{1}{\Gamma(d)\Gamma(e)} \, {}_3 F_2 \bigg[ \begin{matrix} a\ \ b\ \ c \\ d \ \ e\end{matrix};1\bigg]
\ee
to avoid the spurious singularity at $m=p+1$.

\bibliographystyle{JHEP}
\bibliography{bib}
\end{document}